\documentclass[graybox]{svmult}

\usepackage{type1cm}
\usepackage{makeidx}        
\usepackage{graphicx}     
\usepackage{multicol}   
\usepackage[bottom]{footmisc}
\usepackage{newtxtext}
\usepackage{newtxmath} 
\makeindex 

\usepackage{tikz}
\usetikzlibrary{arrows,decorations.pathmorphing,decorations.markings,patterns,decorations.pathreplacing}
\definecolor{BeaverRed}{rgb}{0.8,0,0}
\definecolor{UniWappenBlau}{RGB}{0,0,170}
\usepackage{natbib}
\begin{document}

\title*{Higgs field in cosmology}
\author{Christian F. Steinwachs\\[10mm]
\textit{\noindent Invited contribution to the Heraeus-Seminar ``Hundred Years of Gauge Theory'',\\ 30 July - 3 August 2018, Physikzentrum Bad Honnef,
organized by Silvia De Bianchi and Claus Kiefer. To appear in the proceedings ``100 Years of Gauge Theory. Past, present and future perspectives'' in the series `Fundamental Theories of Physics' (Springer).}
}
\institute{Christian F. Steinwachs \at Department of Physics,  University of Freiburg, Hermann-Herder-Str. 3, 79104 Freiburg, Germany, \email{christian.steinwachs@physik.uni-freiburg.de}}

\authorrunning{Christian F. Steinwachs}
\maketitle

\abstract{The accelerated expansion of the early universe is an integral part of modern cosmology and dynamically realized by the mechanism of inflation. The simplest theoretical description of the inflationary paradigm is based on the assumption of an additional propagating scalar degree of freedom which drives inflation -- the inflaton. In most models of inflation the fundamental nature of the inflaton remains unexplained. 
In the model of Higgs inflation, the inflaton is identified with the Standard Model Higgs boson and connects cosmology with elementary particle physics. A characteristic feature of this model is a non-minimal coupling of the Higgs boson to gravity. I review and discuss several phenomenological and fundamental aspects of this model, including the impact of quantum corrections and the renormalization group, the derivation of initial conditions for Higgs inflation in a quantum cosmological framework and the classical and quantum equivalence of different field parametrizations. }

\section{Introduction}
\label{Intro}
Hermann Weyl's unique style, weaving together physical ideas with mathematical rigour in an encompassing philosophical framework, lead to many important contributions that still play a major role in theoretical physics today. Several aspects discussed  in this contribution are strongly influenced by his ideas on scale invariance, Weyl transformations and on the ``geometrization of physics''. Back in 1918 when the first edition of Raum-Zeit-Materie appeared only three years after Einstein has presented the final form of his theory of General Relativity, Weyl's natural intention was a geometrical unification of gravity and electromagnetism. In this contribution, I discuss a somewhat different kind of unification: the combination of elementary particle physics with cosmology in one unified model. On the one hand, fundamental physics at the electroweak (EW) scale is well described by the Standard Model of particle physics (SM). On the other hand, the cosmic history, including a quasi de Sitter stage of the early Universe, is well described by the Standard Model of Cosmology, which incorporates the mechanism of inflaton. In both models a scalar field plays a central role: the Standard model Higgs boson, which provides the masses to the elementary particles, as well as the cosmological inflaton, which drives the accelerated expansion of the early universe. In this contribution, I discuss the theoretical and observational consequences of the idea that the Higgs boson and the inflaton are one and the same particle.
	
In Sec. \ref{SecInflation} I review the formalism of slow-roll inflation. In Sec. \ref{StandardModel}, I provide the required details about the SM which are relevant for Higgs-inflation.
In Sec. \ref{HiggsInflation} I discuss the model of Higgs inflation. First I show, why a simple identification of the SM Higgs boson with the inflation fails. Then, I demonstrate how a non-minimal coupling of the Higgs boson to gravity can improve the situation. I emphasize the importance of quantum corrections, both from a fundamental as well as from a phenomenological perspective and discuss the aspects related to the renormalization group (RG) improvement of the model. Finally, I show how the initial conditions for the model of non-minimal Higgs-inflation can be derived from quantum cosmological considerations in a self-consistent way. 
In Sec. \ref{QFP}, I discuss the classical and quantum equivalence of different field parametrizations in cosmological models and a field covariant geometric construction of the quantum effective action.
I conclude in Sec. \ref{Conclusions} with a short summary of the discussion.

\section{Inflation}
\label{SecInflation}
In this section, I briefly summarize the essential assumptions and observable consequences of inflation. I discuss the dynamics of the homogeneous and isotropic background and the observational implications following from the linear perturbations propagating on this background.

\subsection{Background dynamics}
Observation support the underlying theoretical assumption that on large scales (${\gtrsim 200}$ Mpc), the universe is spatially flat and statistically homogeneous and isotropic, leading to the Friedman-Lema\^itre-Robertson-Walker (FLRW) line-element
\begin{align}
\mathrm{d}s^2=-\mathrm{d}t^2+a^2(t)\delta_{ij}\mathrm{d}x^{i}\mathrm{d}x^{j}.\label{FLRW}
\end{align}
Here, $t$ and $x^{i}$, $i=1,2,3$ are comoving coordinates and $a(t)$ is the scale factor.
A consistent explanation of the cosmic history requires a \textit{quasi de Sitter stage} in which the early universe expanded accelerated -- inflation. In most theoretical models the inflationary stage is realized dynamically by one or more scalar fields. In contrast to a de Sitter stage corresponding to a cosmological constant $\Lambda$, the dynamics of the scalar inflaton field allows inflation to last sufficiently long to explain the high degree of isotropy of the cosmic microwave background background radiation (CMB), and, at the same time, to end the stage of accelerated expansion, necessary to explain the formation of structure we observe today, see \cite{Starobinsky1979,Starobinsky1980,Guth1981,Mukhanov1981,Mukhanov1982,Starobinsky1982,Linde1982,Albrecht1982,Guth1982,Hawking1982,Bardeen1983}. 
In its simplest incarnation, inflation is described by a single minimally coupled scalar field, the \textit{inflaton} $\varphi$ with action
\begin{align}
S[g,\varphi]=\int\mathrm{d}^4x\sqrt{-g}\left[\frac{M_{\mathrm{P}}^2}{2}R-\frac{1}{2}\partial_{\mu}\varphi\partial^{\mu}\varphi-V(\varphi)\right],
\end{align}
where $M_{\mathrm{P}}=1/\sqrt{8\pi G_{\mathrm{N}}}\approx 2.4\times10^{18}$ GeV (in units $c=\hbar=1$) is the reduced Planck mass, $g=\mathrm{det}(g_{\mu\nu})$ the determinant of the metric field $g_{\mu\nu}$, $R$ the Ricci scalar and $V(\varphi)$ the scalar field potential.
For the symmetric FLRW background \eqref{FLRW} and a homogeneous scalar field $\varphi(t)$,
the field equations for $g_{\mu\nu}$ and $\varphi$ reduce to
\begin{align}
H^2=\frac{1}{3M_{\mathrm{P}}^{2}}\left(\frac{1}{2}{\dot{\varphi}}^2 + V\right),\qquad \ddot{\varphi}+3H\dot{\varphi}+ V_{,\varphi}=0.\label{KG}
\end{align}
Here $H(t):=\dot{a}(t)/a(t)$ is the Hubble parameter and the dot denotes a derivative with respect to Friedman time $t$.
A sufficiently long phase of inflation is realized for a slowly rolling inflaton $\ddot{\varphi}/H^2\ll\dot{\varphi}^2/H^2\ll1$, for which the equations \eqref{KG} reduce to
\begin{align}
H^2\approx \frac{1}{3}\frac{V}{M^{2}_{\mathrm{P}}},\qquad
3H\dot{\varphi}\approx -V_{,\varphi}.\label{SlowRollEOM}
\end{align}
In the slow-roll approximation, the inflationary dynamics is largely dictated by the properties of the inflaton potential $V(\varphi)$. This motivates the introduction of the dimensionless slow-roll parameters which measure the slope and the curvature of $V$, 
\begin{align}
\epsilon_{\mathrm{v}}:=\frac{M_{\mathrm{P}}^2}{2}\,\left(\frac{V_{,\varphi}}{V}\right)^2,\qquad\eta_{\mathrm{v}}:=M_{\mathrm{P}}^2\,\left(\frac{V_{,\varphi\varphi}}{V}\right).\label{SlowRollParameters}
\end{align}
During slow-roll inflation $\epsilon_{\mathrm{v}}\ll1$ and $|\eta_{\mathrm{v}}|\ll1$ quantify the deviation from de Sitter space ($\epsilon_{\mathrm{v}}=\eta_{\mathrm{v}}=0$). More precisely $\varepsilon_{\mathrm{v}}<1$ is required for inflation, i.e. accelerated expansion $\ddot{a}>0$, to be realized in the first place and $|\eta_{\mathrm{v}}|\ll1$ ensures that the inflationary phase lasts sufficiently long to solve the horizon problem. 

\subsection{Perturbations and inflationary observables}
Inflation amplifies the quantized inhomogeneous perturbations around the FLRW background, which provide the seeds for the density perturbations that clump under the influence of gravity and ultimately give rise to the CMB and the structure we observe today. 
The main inflationary observables are the power spectra of the scalar and tensor perturbations, which due to their weak logarithmic $k$-dependence, are parametrized by the power law ansatz
\begin{align}
\mathcal{P}_{\mathrm{t}}:=A_{\mathrm{t}}\,\left(\frac{k}{k_{*}}\right)^{n_{\mathrm{t}}+\ldots},\qquad \mathcal{P}_{\mathrm{s}}:=A_{\mathrm{s}}\,\left(\frac{k}{k_{*}}\right)^{n_{\mathrm{s}}-1+\ldots}\,.\label{PowerSpectra}
\end{align}
The reference scale $k_{*}$ in the observable window ${10^{-4}\text{ Mpc}^{-1}\leq k_{*}\leq10^{-1}\text{ Mpc}^{-1}}$, first crosses the horizon at the moment $k_{*} = a_{*}\, H_{*}$ chosen to correspond to $N=60$, where $N:=\ln a$ is the number of e-folds.
The tensor and scalar amplitudes $A_{\mathrm{t}}$ and $A_{\mathrm{s}}$ characterize the strengths of the power spectra, while the tensor and scalar spectral indices $n_{\mathrm{t}}$ and $n_{\mathrm{s}}$ characterize their tilts, i.e. their weak scale dependence. The ellipsis indicate higher order terms in the expansion, which we have neglect. To first order in the slow-roll approximation, these quantities are assumed to be constant and can be expressed in terms of $V$, $\epsilon_{\mathrm{v}}$ and $\eta_{\mathrm{v}}$,
\begin{align}
A_{\mathrm{t}}={}&\frac{2\,V}{3\,\pi^2\,M_{\mathrm{P}}^4},& A_{\mathrm{s}}={}&\frac{V}{24\,\pi^2\,M_{\mathrm{P}}^4\,\epsilon_{\mathrm{v}}},\label{CMBAmpl}\\
n_{t}={}&-2\,\epsilon_{\mathrm{v}},& n_{\mathrm{s}}={}&1+2\,\eta_{\mathrm{v}}-6\,\epsilon_{\mathrm{v}}\,.\label{Tilt}
\end{align}
All observables \eqref{CMBAmpl} and \eqref{Tilt} are to be evaluated at $\varphi_{*}$, which can be expressed in terms of the number of e-folds $N$ via the integral relation
\begin{align}
N_{*}=\int_{t_{*}}^{t_{\mathrm{end}}}\,\text{d}t\,H\simeq\int^{\varphi_{*}}_{\varphi_{\mathrm{end}}}\,\frac{\text{d}\varphi}{M_{P}^2}\,\frac{V}{V_{,\varphi}}\,.\label{NumberEfolds}
\end{align}
The value $\varphi_{\mathrm{end}}$ is defined by the breakdown of the slow-roll approximation
\begin{align}
\epsilon_{\mathrm{v}}(\varphi_{\mathrm{end}}):=1\,.\label{EndOfInfl}
\end{align}
To first order in the slow-roll approximation, the scalar-to-tensor ratio is given by
\begin{align}
r:=\frac{A_{\mathrm{t}}}{A_{\mathrm{s}}}=16\,\epsilon_{\mathrm{v}}=-8\,n_{\mathrm{t}}\,\label{TTS}.
\end{align}
The last equality is a consistency equation valid in single field models.
There only exists an upper bound on $r$, since the tensor power spectrum has not been measured. Recent observational constraints from the CMB at $k_{*}= 0.05$ $\text{ Mpc}^{-1}$ are provided in \cite{Akrami2018},
\begin{align}
A_{\mathrm{s}}^{*}={}&\left(2.099\pm0.014\right)\times 10^{-9}&& 68\%\;\mathrm{CL},\label{AsCMB}\\
n_{\mathrm{s},\,*} ={}& 0.9649 \pm 0.0042&& 68\%\;\mathrm{CL},\label{Planckns}\\
r_{*} <{}& 0.11&& 95\%\;\mathrm{CL}.\label{Planckr}
\end{align}
For any consistent inflationary model, these observational constraints have to be in agreement with the predictions for \eqref{CMBAmpl} and \eqref{Tilt}.

\section{Standard Model of particle physics}
\label{StandardModel}
The Higgs boson $h$ is an integral part of the SM of particle physics and provides a mechanism by which the SM particles acquire their mass. The Higgs field is a complex $SU(2)$ doublet $\Phi$, which in the real parametrization consists of the radial massive Higgs component $h$ and the three angular Goldstone bosons $\theta_i$, $i=1,2,3$. So far, the Higgs boson is the only fundamental scalar particle which has been detected. It is an interesting question whether there are more (maybe many more) scalar fields in nature. The Higgs sector of the SM is described by the Lagrangian density
\begin{align}
\mathcal{L}_{\mathrm{Higgs}}^{\mathrm{SM}}=-\frac{1}{2}\left|\partial\Phi\right|^2-\frac{\lambda}{4}\left(\left|\Phi\right|^2-v^2\right)^2,\qquad \left|\Phi\right|^2=\Phi^{\dagger}\Phi.
\end{align}
Here $\lambda$ is the quartic Higgs self-coupling and $v\approx 246$ GeV the EW scale. In unitary gauge, with $|\Phi|^2=h^2$, the Lagrangian describing the non-derivative interaction of the Higgs boson $h$ with the other SM particles schematically reads 
\begin{align}
\mathcal{L}_{\mathrm{int}}^{\mathrm{SM}}=-\sum_{\chi}\frac{1}{2}\lambda_{ \chi}\chi^2h^2-\sum_{A}\frac{1}{2}g_A^2A_{\mu}^2h^2-\sum_{\psi}y_{\psi}\bar{\psi}\psi h,
\end{align}
where the sum extends over scalar fields $\chi$, vector gauge fields $A_{\mu}$ and Dirac spinors $\psi$ with the corresponding scalar, gauge and Yukawa couplings $\lambda_{\chi}$, $g_{A}$ and $y_{\psi}$.
The interaction sector is dominated by the heaviest particles in the SM: the Yukawa top-quark, the Higgs boson, the $Z$ boson and the $W^{\pm}$ bosons, with masses 
\begin{align}
M_{\mathrm{t}}^2=\frac{1}{2}y_{\mathrm{t}}^2h^2,\qquad M_{h}^2=2\lambda h^2,\qquad M_{Z}^2=\frac{\left(g+g'\right)^2}{4}h^2,\qquad M_{W}^2=\frac{1}{4}g^2h^2.\label{MassesSM}
\end{align}
Data from collider experiments constrain these masses \cite{Tanabashi2018},
\begin{align}
M_{\mathrm{t}}={}&173.0\pm 0.4\;\mathrm{GeV},&
M_{\mathrm{h}}={}&125.18\pm 0.16\;\mathrm{GeV},\label{Mh}\\
M_{\mathrm{Z}}={}&91.1876\pm 0.0021\;\mathrm{GeV},&
M_{\mathrm{W}}={}&80.379\pm 0.012\;\mathrm{GeV}. \label{MW}
\end{align}
The values for \eqref{MassesSM} and \eqref{Mh} constrain the quartic Higgs self-coupling $\lambda\approx0.1$ at the EW scale $h\simeq v$.

\section{Higgs inflation}
\label{HiggsInflation}
The basic idea of Higgs inflation is to identify the SM Higgs boson $h$ with the cosmic inflaton $\varphi$,
thereby establishing a direct connection between elementary particle physics and inflationary cosmology
\begin{align}
h\equiv\varphi.
\end{align}
Such a unified scenario is not only very appealing from a theoretical point of view, but also very predictive, as it requires to simultaneously match observational constraints from particle physics and cosmology. I first discuss the difference between a minimally and non-minimally coupled SM Higgs to gravity at tree-level and then extend the discussion to the inclusion of important and unavoidable quantum corrections.

\subsection{Minimal Higgs inflation}
\label{sec:1}
The most direct approach to construct the unified scenario of Higgs inflation is to embed the SM in curved spacetime and to analyse the inflationary consequences of a SM Higgs boson minimally coupled to gravity. In this case, the graviton-Higgs sector is described by the action 
\begin{align}
S[g,\varphi]=\int\mathrm{d}^4x\sqrt{-g}\left[\frac{M_{\mathrm{P}}^2}{2}R-\frac{1}{2}\partial_{\mu}\varphi\partial^{\mu}\varphi-\frac{\lambda}{4}\left(\varphi^2-v^2\right)^2\right].\label{MinAct}
\end{align}
The action \eqref{MinAct} was already investigated in one of the earliest models of inflation, formulated in \cite{Linde1983}. Form the viewpoint of the inflationary slow-roll analysis \eqref{MinAct} leads to the chaotic inflation scenario with the monomial potential (for $\varphi/v\gg1$),
\begin{align}
V(\varphi)=\frac{\lambda}{4}\varphi^4.\label{lambdaphi4}
\end{align}
For the inflationary observables \eqref{CMBAmpl}, \eqref{Tilt} and \eqref{TTS} for the potential \eqref{lambdaphi4} to leading order in $N_{*}\gg1$, the model \eqref{MinAct} predicts
\begin{align}
A_{\mathrm{s}}=\frac{2\lambda N_{*}^3}{3\pi^2},\qquad n_\mathrm{s}=1-\frac{3}{N_{*}}=0.95,\qquad r=\frac{16}{N_{*}}\approx 0.2667.\label{InfPredLamb4}
\end{align}
The numerical values are obtained for $N_{*}=60$.
Combining the observational constraint on $A_{\mathrm{s}}$ given in \eqref{AsCMB} with the predicted value in \eqref{InfPredLamb4}, directly translates into a constraint for the quartic Higgs self-coupling $\lambda\approx 10^{-13}$. Such a tiny value for $\lambda$ is clearly incompatible with the value $\lambda\approx0.1$, required by the observational SM constraint on the Higgs mass \eqref{MassesSM} and therefore spoils this first approach to identify the inflaton with the SM Higgs boson.

Note however that the constraint $\lambda\approx0.1$ only has to be satisfied at the EW scale $E_{\mathrm{EW}}\approx 10^2$ GeV, while at inflationary energy scales $E_{\mathrm{inf}}\approx 10^{15}$ GeV, $\lambda$ might attain different values. As I discuss in Sec. \ref{QC}, the RG flow of the SM drives the running $\lambda(t)$ to very small values at high energy scales. But even, if values as small as $\lambda(t_{\mathrm{inf}})\approx 10^{-13}$ could be attained dynamically by the RG flow, such that the CMB normalization condition \eqref{AsCMB} would be satisfied at the energy scale of inflation, the RG corrections would also have to improve the situation with the spectral observables as the tree-level chaotic inflationary model with potential \eqref{lambdaphi4} predicts a scalar spectral index and a tensor-to-scalar ratio \eqref{InfPredLamb4} incompatible with the observational constraints \eqref{Planckns} and \eqref{Planckr}. 

\subsection{Non-minimal Higgs-inflation}
The central assumption for a successful identification of the inflaton with the SM Higgs boson is to include a non-minimally coupling of $\varphi$ to gravity.

Early ideas to incorporate a non-minimal coupling of an abstract inflaton field to gravity were formulated in \cite{Salopek1989,Fakir1990} in order to improve the situation with the observational constraints on $\lambda$ and on the spectral observables \eqref{InfPredLamb4}. The identification of a non-minimally coupled inflaton with the SM Higgs boson was proposed in \cite{Bezrukov2008}.
Independent of its phenomenological impact on the  observational constraints, a non-minimal coupling might be motivated by several theoretical reasons: First, a non-minimal coupling might, to some extend, be viewed as incorporating the Machian idea of a variational gravitational constant. Second, its presence is required for technical reasons. Even in the absence of a non-minimal coupling, already the first quantum corrections for a self-interacting scalar field induce a non-minimal coupling term and the consistency of the renormalization procedure requires that this term must be included in the action. Third, from an effective field theory point of view, the non-minimal coupling term in the action corresponds to a marginal operator, which is on equal footing with the Einstein-Hilbert term in a derivative expansion and should therefore be included in the defining low energy limit of the theory. Fourth, the inclusion of such a term leads to an asymptotic scale invariance for large values of the scalar field, which realizes inflation in a natural way and ends it when the scale invariance is explicitly broken by the Einstein-Hilbert operator. Fifth, in the context of effective string theory inspired models, a non-minimal coupling unavoidably arises in the form of a dilaton or moduli field. Irrespectively of these theoretical motivations, in the following sections, I discuss the phenomenological consequences of a non-minimal coupling in the context of Higgs inflation.

\begin{align}
S[g,\varphi]=\int\mathrm{d}^4x\sqrt{-g}\left[\frac{1}{2}\left(M_{\mathrm{P}}^2+\xi\varphi^2\right)R-\frac{1}{2}\partial_{\mu}\varphi\partial^{\mu}\varphi-\frac{\lambda}{4}\left(\varphi^2-v^2\right)^2\right].\label{JFAct}
\end{align}
The formalism of Sec.~\ref{SecInflation} is directly applicable by performing a field redefinition 
\begin{align}
\hat{g}_{\mu\nu}=\left(1+\xi\frac{\varphi^2}{M_{\mathrm{P}}^2}\right)g_{\mu\nu},\qquad
\left(\frac{\partial\hat{\varphi}}{\partial\varphi}\right)=\frac{1+\xi\left(1+6\xi\right)\frac{\varphi^2}{M_{\mathrm{P}}^2}}{\left(1+\xi\frac{\varphi^2}{M_{\mathrm{P}}^2}\right)^2}.\label{TrafoJFEF}
\end{align}
In this way, the  action \eqref{JFAct}, originally formulated in the Jordan frame (JF) variables $(g_{\mu\nu},\varphi)$, is mapped to the action in the Einstein frame (EF) variables $(\hat{g}_{\mu\nu},\hat{\varphi})$,\footnote{Since the form of the EF action \eqref{EFACT} formally resembles that of the Einstein-Hilbert action \eqref{MinAct}, the formulation in terms of the variables $(\hat{g}_{\mu\nu},\hat{\varphi})$ is called ``Einstein frame''. The formulation in terms of the ``Jordan frame'' variables $(g_{\mu\nu},\varphi)$, for which the scalar-tensor character of the action is manifest, derives from the early work of Pascual Jordan on such models.}
\begin{align}
\hat{S}[\hat{g},\hat{\varphi}]=\int\mathrm{d}^4x\sqrt{-\hat{g}}\left[\frac{1}{2}M_{\mathrm{P}}^2\hat{R}-\frac{1}{2}\hat{g}^{\mu\nu}\partial_{\mu}\hat{\varphi}\partial_{\nu}\hat{\varphi}-\hat{V}(\hat{\varphi})\right],\label{EFACT}
\end{align}
with the EF potential defined as
\begin{align}
\hat{V}(\hat{\varphi}):=\frac{M_{\mathrm{P}}^4\lambda}{4\xi^2}\left(1-e^{-\sqrt{\frac{2}{3}}\frac{\hat{\varphi}}{M_{\mathrm{P}}}}\right)^2.\label{EFPot}
\end{align}
The main purpose of this transformation is to remove the non-minimal coupling by a Weyl transformation of the metric field $g_{\mu\nu}\to \hat{g}_{\mu\nu}=\Omega^2(\varphi)g_{\mu\nu}$, with a field-dependent conformal factor $\Omega^2(\varphi)$ explicitly given in the first equation of \eqref{TrafoJFEF}. Since the derivatives of $\Omega(\varphi)$, which arise in this transformation, also induce a contribution to the kinetic term of the scalar field, a reparametrization $\varphi\to\hat{\varphi}$ is required to obtain a canonically normalized kinetic term. Ultimately, in this way the complexity associated with the non-minimal coupling is shifted to the scalar potential. While the EF action formally resembles that of the minimally coupled scalar field, the potentials differ and matter fields also feel the coupling to the scalar-field dependent EF metric.\footnote{Note however, that matter fields  which do not directly couple to $\varphi$ and whose action is invariant under Weyl transformations are insensitive to the transformation \eqref{TrafoJFEF} -- at least at the classical level.}

Applying the inflationary formalism of Sec.~\ref{SecInflation} to the EF action \eqref{EFACT}, the inflationary observables for the potential \eqref{EFPot}  to leading order in $\hat{N}_{*}\gg1$ are
\begin{align}
\hat{A}_{\mathrm{s}}^{*}=\frac{\hat{N}_{*}^2}{72 \pi^2 }\frac{\lambda }{\xi^2},\qquad \hat{n}_{\mathrm{s}}^{*}=1-\frac{2}{\hat{N}_{*}}\approx0.9667,\qquad \hat{r}^{*}=\frac{12}{\hat{N}_{*}^2}\approx 0.0033.\label{ObsNMHI}
\end{align}
The numerical values are again presented for $N_{*}=60$.
Comparing the potential \eqref{lambdaphi4} of the minimally coupled action \eqref{MinAct} with the EF potential \eqref{EFPot}, there are two main effects which improve the situation with the observational constraints:
First, for the non-minimally coupled Higgs boson, the normalization of the EF potential depends on the ratio $\lambda/\xi^2$ in contrast to the pure $\lambda$ dependence for the minimally coupled Higgs boson. Therefore, by making $\xi$ sufficiently large $\xi\approx10^{4}$, the quartic coupling can be tuned to $\lambda\approx0.1$ such that both the CMB constraint \eqref{AsCMB} as well as the Higgs mass constraint \eqref{Mh} are satisfied simultaneously.
The second effect is that, in contrast to the quartic chaotic inflation potential \eqref{lambdaphi4}, the EF potential \eqref{EFPot} becomes field independent and almost flat for large field values of $\hat{\varphi}$, thereby improving the situation with the constraints on the spectral observables in \eqref{ObsNMHI} -- both the spectral index as well as the tensor-to-scalar ratio are in perfect agreement with the observational bounds \eqref{Planckns} and \eqref{Planckr}.

\subsection{Quantum corrections and the renormalization group}
\label{QC}
When identifying the non-minimally coupled SM Higgs boson with the inflaton at tree-level, the main compatibility requirement is that the observational restrictions on the CMB normalization \eqref{AsCMB} and the Higgs mass \eqref{Mh} are satisfied simultaneously, requiring $\xi\approx10^{4}$ and $\lambda\approx10^{-1}$.
As can be seen from \eqref{ObsNMHI}, the scalar-tensor sector of the non-minimally coupled Higgs field belongs to a more general class of inflationary models, for which the predictions of the scalar spectral index and the tensor-to-scalar ratio are independent of the model parameters. Therefore, provided the constraints \eqref{AsCMB} and \eqref{Mh} are satisfied, the inflationary predictions of the tree-level non-minimal Higgs inflation model are insensitive to the detailed properties of the SM particles.

It is clear that such a tree-level consideration is incomplete and cannot be correct. The quantum loop corrections of the SM particles have to be taken into account.
In particular, the quantum corrections to the EF effective potential, which are dominated by the heaviest particles of the SM, ultimately induce a dependence of the spectral observables $n_{\mathrm{s}}$ and $r$ on the particle content of the SM. In \cite{Barvinsky2008}, the one-loop effective  Coleman-Weinberg potential in the EF (expressed in terms of the JF field $\varphi$) was obtained as
\begin{align}
\hat{V}=\frac{\lambda M_{\mathrm{P}}^4}{4\xi^2}\left(1-\frac{2M_{\mathrm{P}}^2}{\xi\varphi^2}+\frac{\mathbf{A}_{\mathrm{I}}}{16\pi^2}\ln\frac{\varphi}{\mu_0}\right),
\label{EffPotEF}
\end{align}
with arbitrary renormalization point $\mu_0$ and the inflationary anomalous scaling
\begin{align}
\mathbf{A}_{\mathrm{I}}=\mathbf{A}-12\lambda=\frac{3}{8\lambda}\left[2g^4+\left(g^2+g'^2\right)^2-y_{\mathrm{t}}^4\right]-6\lambda.\label{InfAnDim}
\end{align}
Here $g$ and $g'$ are the EW gauge couplings and $y_{\mathrm{t}}$ the Yukawa to-quark coupling.
As shown in \cite{Barvinsky2008}, the impact of these quantum corrections lead to essential modifications of the shape of the inflationary potential, as during inflation $\varphi\gg M_{\mathrm{P}}/\sqrt{\xi}$, the second term in \eqref{EffPotEF} is negligible and the logarithmic quantum corrections dominate over the flat tree-level part.
The quantum contribution can be parametrized by the dimensionless quantity
\begin{align}
x:=\frac{\hat{N} \mathbf{A}_{I}}{48\pi^2},
\end{align}
which enters the inflationary observables \eqref{ObsNMHI} in the form of correction factors
\begin{align}
\hat{A}_{\mathrm{s}}^{*}=\frac{\hat{N}_{*}^2}{72 \pi^2 }\frac{\lambda }{\xi^2}\left(\frac{e^{x}-1}{xe^x}\right)^2,\qquad \hat{n}_{\mathrm{s}}^{*}=1-\frac{2}{\hat{N}_{*}}\frac{x}{e^x-1},\qquad \hat{r}^{*}=\frac{12}{\hat{N}_{*}^2}\left(\frac{xe^x}{e^x-1}\right).\label{ObsNMHIQC}
\end{align}
As demonstrated in \cite{Barvinsky2008}, the impact of these quantum corrections could render the identification of the SM Higgs with the inflaton invalid. 

However, taking into account the quantum corrections at the EW scale is not sufficient. The coupling constants depend on the energy scale of the underlying physical process and their change is determined by the system RG equations
\begin{align}
\frac{\mathrm{d}g_{i}}{\mathrm{d}t}=\beta_{i},\qquad\frac{\mathrm{d}Z}{\mathrm{d}t}=\gamma Z.
\end{align}
The beta functions $\beta_i$ of the couplings $g_i=\{\lambda,\,y_{\mathrm{t}},\,\xi,\,g,\,g',\,g_{\mathrm{s}}\}$, include the quartic Higgs self-coupling $\lambda$, the Yukawa top-quark coupling $y_t$, the non-minimal coupling $\xi$, the EW gauge couplings $g$, $g'$ and the strong gauge coupling $g_{\mathrm{s}}$ all depend on the logarithmic RG scale ${t=\ln\left(\varphi/M_{\mathrm{t}}\right)}$ where the arbitrary renormalization point $\mu_0$ has ben fixed to agree with the highest mass scale $M_{\mathrm{t}}$ in the SM. In addition, the wave function renormalization $Z$ of the Higgs boson determined by its anomalous dimension $\gamma$ have to be taken into account. However, since it was found in \cite{Barvinsky2009} that the running of $Z$ is very slow, we neglect the running of $Z$ and assume $Z=1$ in the following discussion.
Although, the dimensionless couplings $g_i$, only change logarithmically with the energy scale, the effects of the RG improvement becomes sizeable as the EW scale and the inflationary scale are separated by approximately thirteen orders of magnitude.  
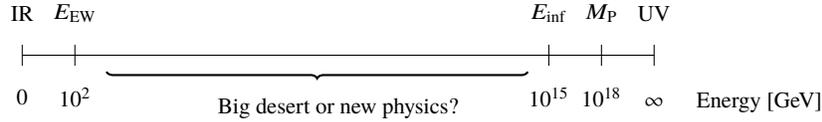
\begin{figure}[h!]
\begin{center}
\begin{tikzpicture}[scale=1.4]
\begin{scope}
\draw[](-1,0)--(5,0);
\node at (6,-0.45){\small Energy [GeV]};
\draw (5,-0.1)--(5,0.1);
\node at (5,0.4) {UV};
\node at (5,-0.45) {$\infty$};

\draw (4,-0.1)--(4,0.1);
\node at (4,0.4){$E_{\mathrm{inf}}$};
\node at (4,-0.4){$10^{15}$};

\draw (4.5,-0.1)--(4.5,0.1);
\node at (4.5,0.4){$M_{\mathrm{P}}$};
\node at (4.5,-0.4){$10^{18}$};

\draw (-0.5,-0.1)--(-0.5,0.1);
\node at (-0.5,0.4){$E_{\mathrm{EW}}$};
\node at (-0.5,-0.4){$10^{2}$};

\draw (-1,-0.1)--(-1,0.1);
\node at (-1,0.4){IR};
\node at (-1,-0.4){$0$};
\draw [thick,decoration={brace,mirror,raise=0.5cm},decorate](-0.2,0.2)--(3.8,0.2);
\node at (2,-0.5){Big desert or new physics?};
\end{scope}
\end{tikzpicture}	
\end{center}
\caption{Different energy scales and their connection to the model of non-minimal Higgs inflation.}
\end{figure}

\noindent The dominant one-loop RG improvement of the model was investigated in \cite{Bezrukov2009a,DeSimone2009,Barvinsky2009} and shown to be the essential mechanism by which the Higgs inflation compatibility constraints can be satisfied. The subleading two-loop contributions to the running, first considered in \cite{Bezrukov2009}, are significant and, compared to the one-loop running, reduce the bound of the cosmologically compatible Higgs mass about $10$ GeV down to the observed value \eqref{Mh}.\footnote{A non-perturbative treatment in the context of the asymptotic safety paradigm which includes the running of Newtons constant has been investigated in \cite{Shaposhnikov2010}. Imposing that $\lambda$ and $\beta_{\lambda}$ vanish at the Planck scale and evolving the flow towards the IR leads to a Higgs mass prediction of $M_{\mathrm{h}}\approx126$ GeV.}

In general, the functional shape of the tree-level EF potential \eqref{EFPot} is changed by the RG improvement, which in turn leads to modified predictions for the inflationary observables \eqref{ObsNMHI}. Moreover, since the RG flow of the SM is very sensitive to the initial conditions of $\lambda(t)$ and $y_{\mathrm{t}}(t)$ at the EW scale $t_{\mathrm{EW}}\approx 0$, and since these initial conditions are related to the observed masses $M_{\mathrm{h}}$ and $M_{\mathrm{t}}$, the RG improved spectral observables can induce a strong dependence on the precise values of these masses and therefore on the details of the SM.

The non-minimal coupling  in the JF directly couples the Higgs boson $h$ to derivatives of the metric field and thereby mixes gravitational and Higgs degrees of freedom. In contrast, the field content of the  graviton-scalar sector in the EF is diagonal and leads to a Higgs propagator
\begin{align}
\langle h(x),h(0)\rangle
\propto\frac{s(\varphi)}{\hat{g}^{1/2}\left(\hat{\Box}-M_{\mathrm{h}}^2\right)}.\label{HiggsEFprop}
\end{align}
In \cite{DeSimone2009,Barvinsky2009}, it was demonstrated that the impact of the non-minimal coupling on the SM beta functions can be effectively incorporated by taking into account the weighting of the internal Higgs propagators in the corresponding Feynman diagrams by a power of the suppression function
\begin{align}
s(\varphi)=\frac{U}{G U+3(U_1)^2}\simeq
\begin{cases}
1&\text{ for }\varphi\ll \frac{M_{\mathrm{P}}}{\sqrt{\xi}},\\
\frac{1}{6\xi}&\text{ for }\varphi\gg \frac{M_{\mathrm{P}}}{\sqrt{\xi}}.
\end{cases}\label{supps}
\end{align}
Here the functions $U(\varphi)$ and $G(\varphi)$ are defined in \eqref{ActST}, and, upon comparison with the model of non-minimal Higgs inflation \eqref{JFAct}, are given by $U=(M_{\mathrm{P}}^2+\xi\varphi^2)/2$ and $G(\varphi)=1$. Since the suppression function \eqref{supps} behaves like $s\approx 1/6\xi$ for large field values, a large non-minimal coupling ${\xi\approx 10^{4}}$ leads to a strong suppression of Higgs contributions in the beta functions. Prior to the Higgs discovery, the Higgs mass was expected to lie in the interval ${118\;\mathrm{ GeV}\lesssim M_{\mathrm{h}}\lesssim 180\;\mathrm{ GeV}}$ and the suppression phenomenon was termed ``asymptotic freedom''  in \cite{Barvinsky2009}, as the suppression of Higgs contribution in the beta function $\beta_{\lambda}$ essentially prevents $\lambda(t)$ to run into a Landau pole before the energy scale of inflation and allows a perturbative treatment up to inflationary energy scales -- even for large Higgs mass values. 
\begin{figure}
	\begin{center}
		\includegraphics[width=5cm,height=3cm]{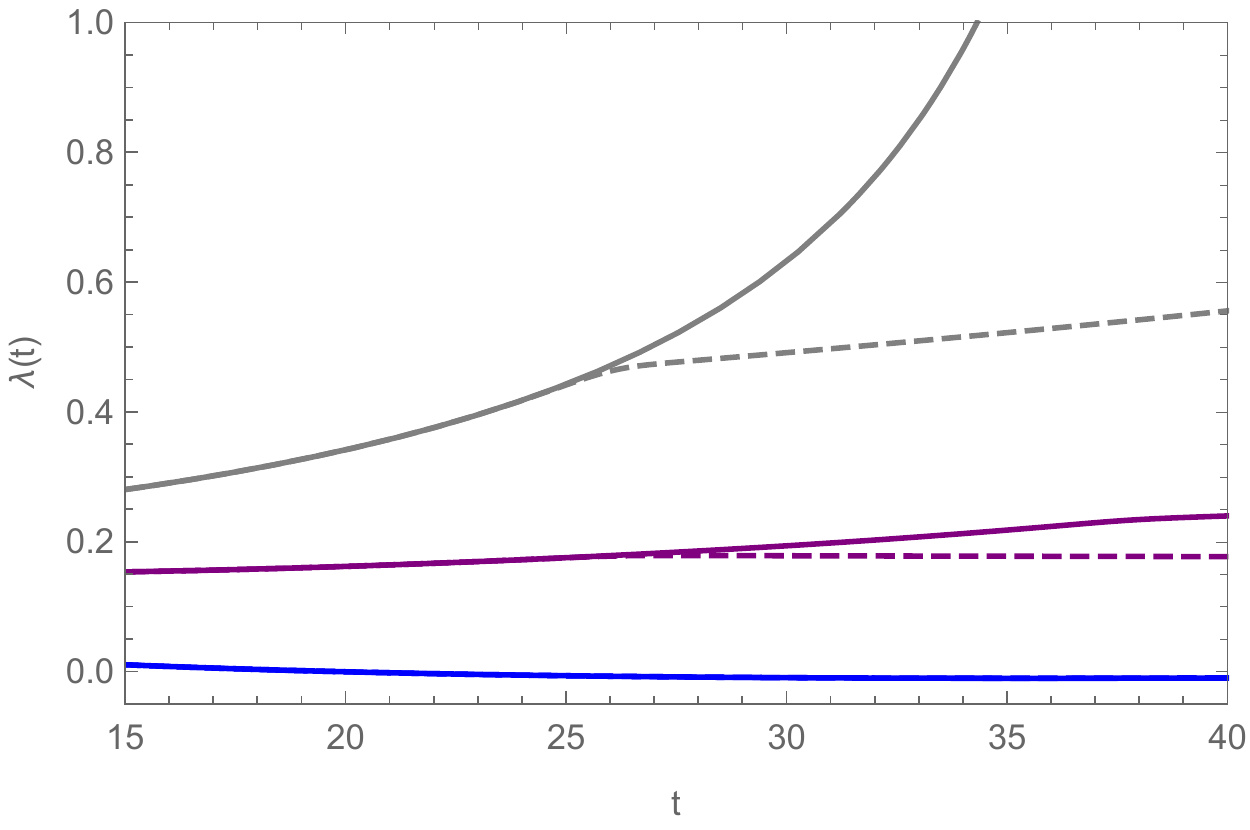}
	\end{center}
	\caption{The RG running of $\lambda(t)$ for fixed top-quark mass $M_{\mathrm{t}}=173$ GeV and different Higgs masses $M_{\mathrm{h}}=180$ GeV (grey), $M_{\mathrm{h}}=160$ GeV (purple) and $M_{\mathrm{h}}=125$ GeV (blue) including the effect of the non-minimal coupling due to the propagator weighting with the suppression function (dashed) and without the propagator weighting (solid). For large Higgs masses, the unsuppressed beta functions would drive $\lambda(t)$ into a Landau pole for scales below the energy scale of inflation $t<t_{\mathrm{inf}}$ (grey solid line). The numerically integrated two-loop beta functions with the weighting were taken from \cite{Allison2014} and those without weighting from the two-loop truncation of the beta functions in the appendix of \cite{Buttazzo2013}.}	
	\label{FigSupp}
\end{figure}

\noindent From Fig. \ref{FigSupp}, it is clear that the suppression is strongest for large Higgs masses. Since the discovery of a light Higgs with $M_{\mathrm{h}}\approx125$ GeV, it is clear that the suppression mechanism is no longer very relevant.\footnote{It might however become relevant for strong non-minimal couplings $\xi\gg10^{4}$ arising e.g. in induced inflation -- not because the stronger suppression with $s=1/6\xi$, but because transition regime $M_{\mathrm{P}}/\xi\lesssim \varphi\lesssim M_{\mathrm{P}}/\sqrt{\xi}$ during which the suppression mechanism becomes effective is essentially lowered for large $\xi$ and could therefore lead to a strong suppression of $\lambda$-dependent terms at a stage where the RG flow has not driven $\lambda$ to values $\lambda\ll0.1$.} Instead, the RG flow of the SM drives the running $\lambda(t)$ to very small values at high energy scales and $\lambda$-dependent contributions in the beta functions are anyway small. Therefore, the influence of the non-minimal coupling $\xi$ on the SM beta functions is very weak. Moreover, given that the running of the non-minimal coupling $\xi$ itself is rather slow, we neglect its running $\beta_{\xi}\approx0$ and consider only the system of unmodified pure SM beta functions in the following discussion.
\begin{figure}
	\begin{center}
		\includegraphics[width=5cm,height=3cm]{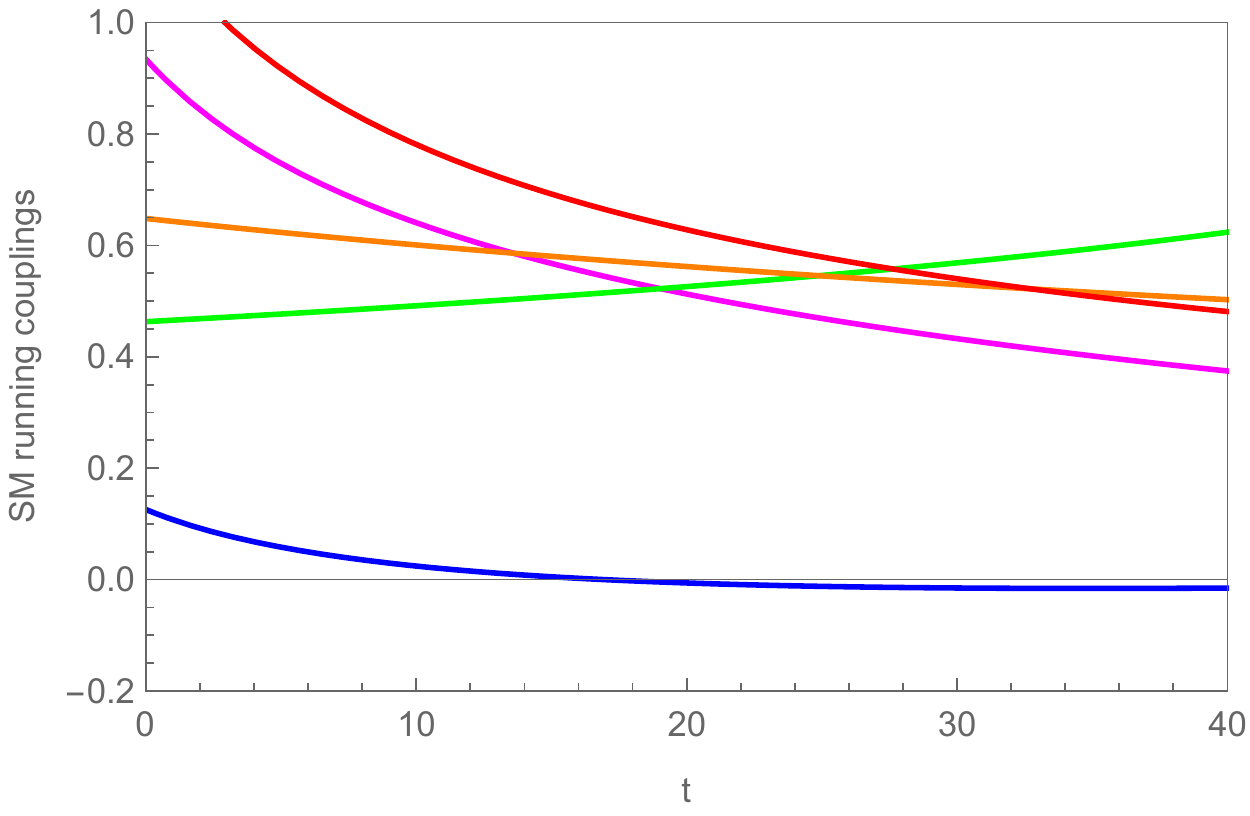}\qquad\qquad
		\includegraphics[width=5cm,height=3cm]{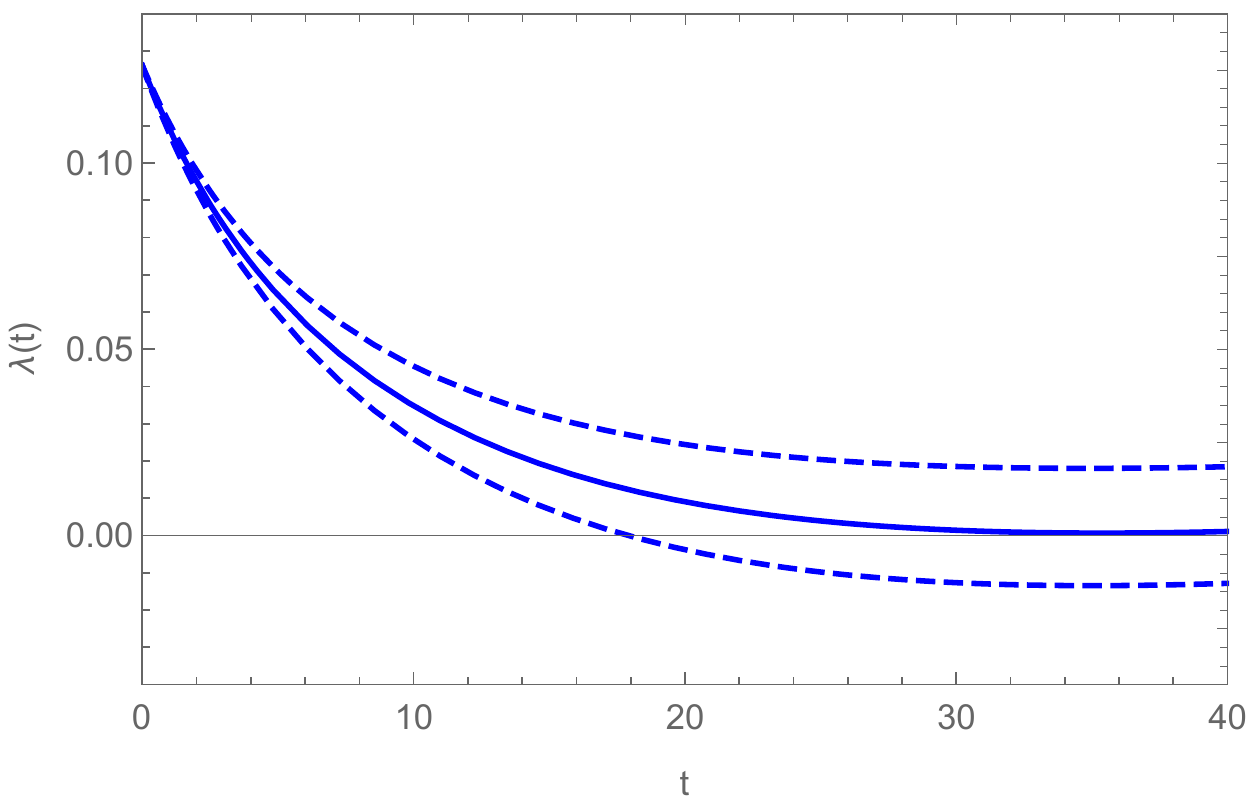}
	\end{center}
	\caption{Left: The pure SM running of the quartic Higgs self-coupling $\lambda$ (blue), the Yukawa top quark coupling $y_t$ (pink) and the EW and strong gauge couplings $g$ (green line), $g'$ (orange line) and $g_{\mathrm{s}}$ (red line). Right: Zoomed in plot of the pure SM running of the quartic Higgs self-coupling $\lambda$ for fixed Higgs mass $M_{\mathrm{h}}=125$ GeV and three different values of the Yukawa top quark mass $M_{\mathrm{t}}=168$ GeV (upper blue dashed line), $M_{\mathrm{t}}=173$ GeV (middle blue solid line), and $M_{\mathrm{t}}=170.8$ GeV (lower blue dashed line), illustrating the the Higgs coupling might be driven to negative values, depending on the precise value of $M_{\mathrm{t}}$. In both plots the two-loop approximation of the beta functions presented in \cite{Buttazzo2013} were numerically integrated with \texttt{Mathematica}.}
	\label{FigSMRunning}	
\end{figure}

\noindent Even if the running of the non-minimal coupling and its impact on the running of the SM beta functions is rather mild for a light SM Higgs boson, the presence of the non-minimal coupling is nevertheless crucial, as it ensures that the EF potential \eqref{EFPot} is almost field independent (i.e. flat) with the overall normalization factor $\lambda(t)/\xi^2$. Before I address the (in)stability of the RG improved EF potential, I briefly discuss three qualitatively different scenarios under the assumption that $\lambda(t)>0$ for all $t$.    
For a light Higgs boson, the RG flow of the SM drives $\lambda(t)$ to very small values at high energies. Moreover, its flow is also very slow at high energies $\beta_{\lambda}\ll1$ and $\lambda(t)$ develops a minimum $\lambda_0:=\lambda(t_0)$ at $t_0$ defined by $\partial_{t}\lambda(t)|_{t=t_0}=\beta_\lambda(t)|_{t=t_0}=0$. Hence, the running in the vicinity of $\lambda_0$ might be described by the Taylor expansion
\begin{align}
\lambda(t)=\lambda_0+\frac{\lambda_2}{(16\pi^2)^2}\, t^2+\mathcal{O}(t^3).
\end{align} 
In general, the values of $\lambda_0$ and $\lambda_2$ are functions of the SM input and depend predominately on the values of $M_{\mathrm{h}}$ and $M_{\mathrm{t}}$. Numerical integration of the RG flow reveals however that $\lambda_2$ is rather insensitive to changes in $M_{\mathrm{h}}$ and $M_{\mathrm{t}}$ and is well approximated by a constant $\lambda_2/(16\pi^2)^2\approx4\times10^{-5}$. In contrast, the value of $\lambda_0$ varies between $10^{-2}$ and $10^{-6}$ and its dependence on $M_{\mathrm{h}}$ and $M_{\mathrm{t}}$ can be parametrized by a fitting formula as e.g. discussed in \cite{Hamada2014,Bezrukov2014}; see also the discussion in the recent review article \cite{Rubio2019}.
The value $t_{0}$ is related to a value $t_{\mathrm{crit}}$ at which the RG improved effective EF potential has an inflection point  $\partial_t^2\hat{V}(t)|_{t=t_{\mathrm{crit}}}:=0$. There are three qualitatively different scenarios, depending on the sign of the slope $\partial_t\hat{V}(t)|_{t=t_{\mathrm{crit}}}$ at that point :
\begin{enumerate}
	\item[I.] {\bf Universal: $\partial_t\hat{V}(t)|_{t=t_{\mathrm{crit}}}\gg0$} \\
	\begin{figure}
		\begin{center}
			\includegraphics[width=11.5cm,height=3cm]{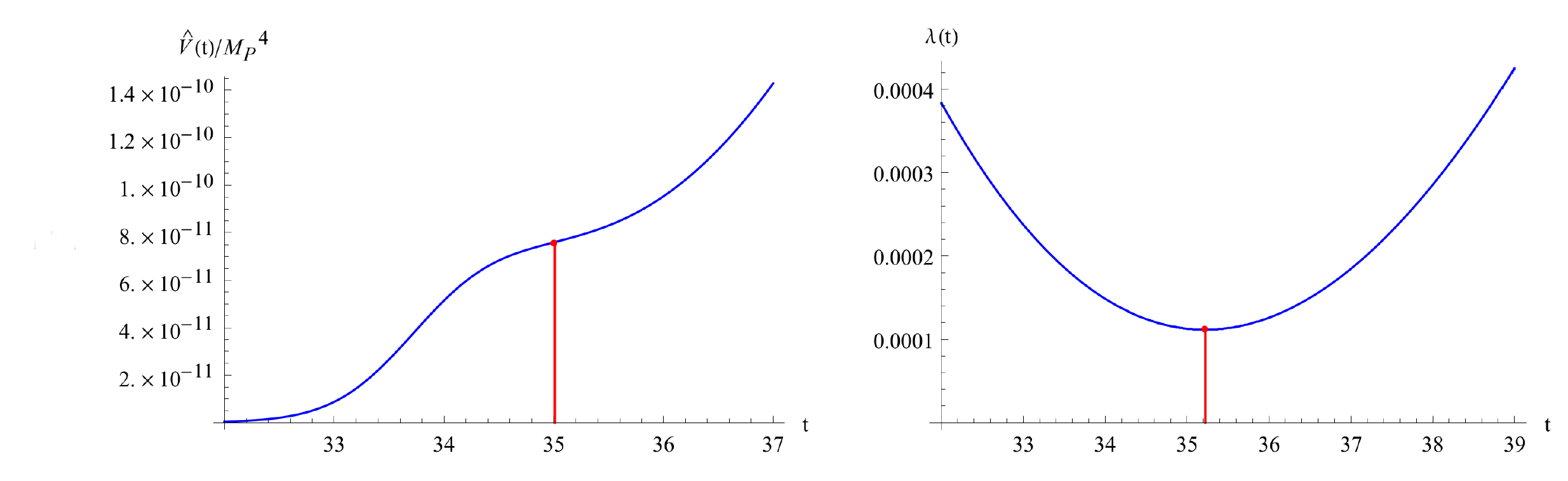}
		\end{center}
		\caption{Left: The RG improved EF potential is a monotonically increasing function. The red line connects $t_{\mathrm{crit}}$ with $\partial_t^2\hat{V}(t)|_{t=t_{\mathrm{crit}}}$. Right: The running of $\lambda$, where the red line connects $t_{0}$ with $\lambda_0$.}
		\label{FigUniv}
	\end{figure}

\noindent The positive slope of the RG improved EF potential cannot be too large (such as exaggeratedly shown in Fig. \ref{FigUniv} for illustrative purposes) and must be cut off by a sufficiently strong non-minimal coupling $\xi$ in order not to spoil the flatness of the potential required for slow-roll inflation. Since the shape of the RG improved potential is almost unchanged compared to the shape of the tree-level EF potential, the RG improved spectral observables \eqref{ObsNMHI}, which depend on derivatives of the potential, are almost identical to the tree-level predictions \eqref{EFPot} in this scenario. Therefore the cosmological predictions are largely insensitive to the details of the SM in this scenario. \\[2mm]

	\item[II.]{\bf Critical: $\partial_t\hat{V}(t)|_{t=t_{\mathrm{crit}}}\geq0$} \\
	\begin{figure}[h!]
		\begin{center}
			\includegraphics[width=11.5cm,height=3cm]{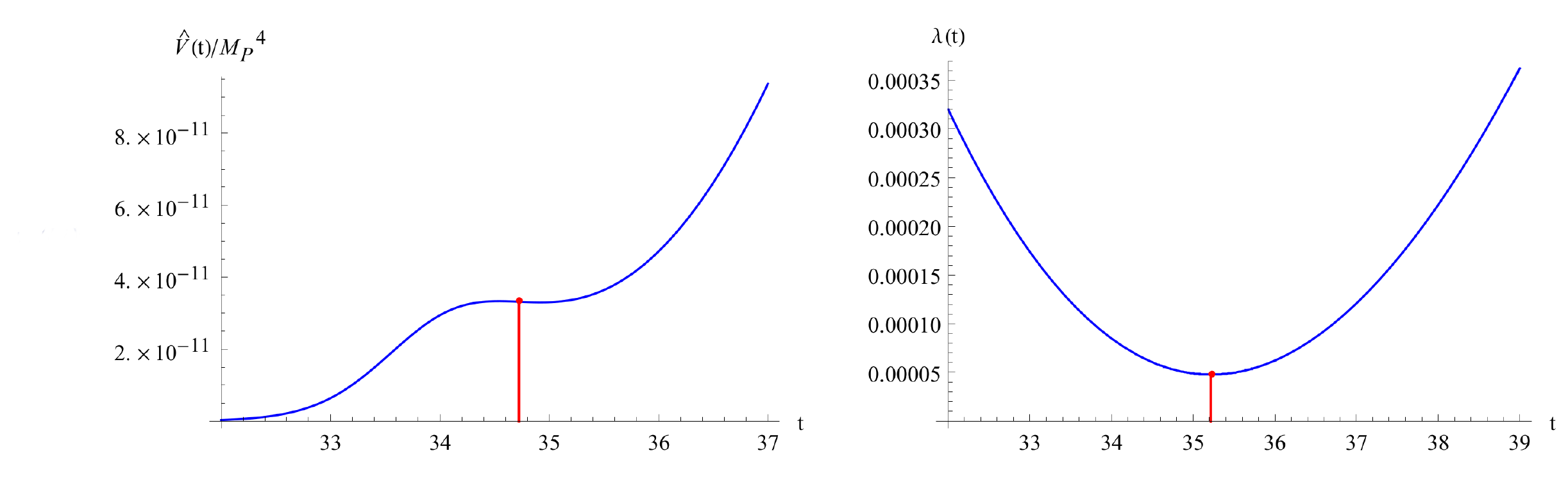}
		\end{center}
	\caption{Left: The inflection point of the RG improved EF potential coincides with its extremum. The red line connects connects $t_{\mathrm{crit}}$ with $\partial_t^2\hat{V}(t)|_{t=t_{\mathrm{crit}}}$. Left: The RG running of $\lambda(t)$. The red line connects $t_0$ with the minimum $\lambda_0$. }
	\end{figure}

\noindent The exact relation $\partial_t\hat{V}(t)|_{t=t_{\mathrm{crit}}}=0$ would lead to a strictly constant plateau of the RG improved EF potential, preventing any inflationary dynamics. In contrast a slight violation of this strict condition by allowing $\partial_t\hat{V}(t)|_{t=t_{\mathrm{crit}}}\geq0$ with $\left|\partial_t\hat{V}(t)|_{t=t_{\mathrm{crit}}}\right|\ll1$ leads to an extremely flat plateau on which (ultra) slow-roll inflation can take place. 
Since this configuration can only be obtained by a highly fine-tuned combination of parameters $M_{\mathrm{h}}$, $M_{\mathrm{t}}$ and $\xi$, the RG improved cosmological predictions of non-minimal Higgs inflation in the critical regime strongly depend on the details of the SM at the EW scale, in particular on the values of $M_{\mathrm{t}}$. 
Compared to its value at the EW scale $\lambda_{\mathrm{EW}}\approx0.1$, the running $\lambda$ can be as small as $\lambda_{\mathrm{inf}}\approx 10^{-6}$ during inflation, such that the CMB normalization condition \eqref{ObsNMHI} for $A_{\mathrm{s}}$ allows for a significant smaller $\xi=\mathcal{O}(10)$ as found in \cite{Allison2014,Bezrukov2014,Hamada2014}.

During the (ultra) slow-roll dynamics on the (ultra) flat plateau, the background dynamics of the inflaton field might no longer be dominated by the overall classical slow-roll drift but by quantum fluctuations, a scenario which can be consistently described within the stochastic approach, see \cite{Starobinsky1994}. In contrast to the universal regime, the slow-roll parameter $\varepsilon_{\mathrm{v}}(\varphi)$ defined in \eqref{SlowRollParameters} is no longer a monotonic function of $\varphi$, but changes the sign of its slope in accordance with the change of slope of the RG improved EF potential at the inflection point. In particular, the tree-level consistency condition \eqref{TTS} implies that, in contrast to the small universal tree-level prediction $r\approx 3\times 10^{-3}$,  in the critical scenario the tensor-to-scalar ratio $r$ can attain larger values up to $r=\mathcal{O}(10^{-1})$, c.f.  \cite{Allison2014,Bezrukov2014,Hamada2014}. In general, the non-monotonic behaviour of the slow-roll parameter $\varepsilon_{\mathrm{v}}$ also leads to a rather strong change ($k$-dependence) of $n_{\mathrm{s}}$ where the simple power-law parametrization of the primordial power spectra \eqref{PowerSpectra} is no longer appropriate.\\[3mm]

\item[III.]{\bf Hilltop: $\partial_t\hat{V}(t)|_{t=t_{\mathrm{crit}}}<0$}\\ 
\begin{figure}
	\begin{center}
		\includegraphics[width=11.5cm,height=3cm]{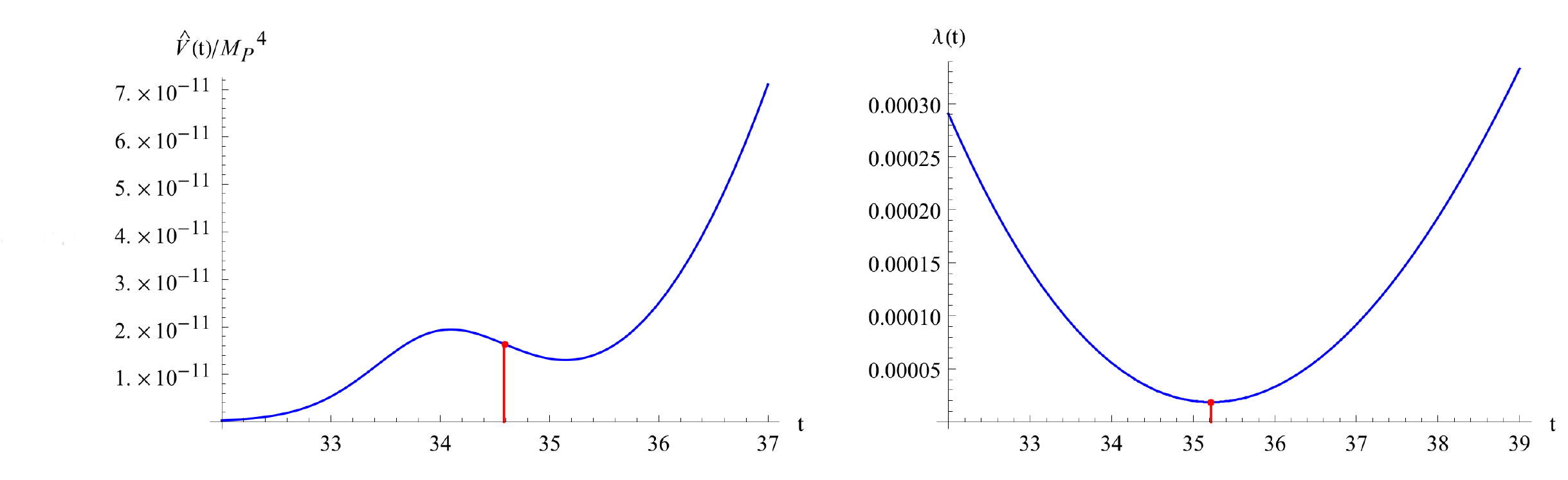}
	\end{center}
\caption{Left: The RG improved EF potential forms a second minimum at high energies.  The red lie connects $t_{\mathrm{crit}}$ with the inflection point $\partial_t^2\hat{V}(t)|_{t=t_{\mathrm{crit}}}$. Left: The RG running of $\lambda(t)$. The red line connects $t_0$ with the minimum $\lambda_0$.}
\end{figure}

\noindent The RG improved EF potential develops a second local minimum at high energy scales. By lowering the values of $\lambda_0$ (increasing $M_{\mathrm{t}}$ for fixed $M_{\mathrm{h}}$), the second local minimum can be continuously lowered, up to the point where $\lambda_0=0$ and its height degenerates with the EW vacuum.
The two minima are separated by a local maximum at which hilltop inflation can take place. Such a behaviour of the RG improved EF potential has been found in \cite{Barvinsky2009}, see also \cite{Enckell2018a}. In order to realize a successful phase of inflation, it must be ensured that the inflaton field can roll down all the way to the EW vacuum and does not get trapped in the second (false) vacuum when rolling down the hilltop in the opposite direction. In general, this scenario would require a rather strong fine tuning to arrange for the correct initial conditions of inflation. However, as I discuss in Sec. \ref{SecQC}, the formation of the initial conditions for non-minimal Higgs inflation might be consistently derived from more fundamental quantum cosmological considerations. 
\end{enumerate}

\subsection{Instability of the electroweak vacuum}
The stability of the EW vacuum together with the associated restrictions on the SM masses has been investigated already in \cite{Arnold1989,Sher1989,Anderson1990,Espinosa1995, Froggatt1996,Isidori2008}. The RG flow of the SM is known to high precision and for the central values of the Higgs mass and the top-quark mass at the EW scale \eqref{Mh}, the RG flow of the SM drives $\lambda$ to negative values $\lambda(t)<0$ at high energies $t_{\mathrm{inst}}<t<t_{\mathrm{inf}}$, see \cite{Degrassi2012,Bezrukov2012a,Buttazzo2013,Bednyakov2015}.
A negative $\lambda$ leads to the formation of a negative global minimum at an energy scales below the energy scale of inflation and therefore to a non-stable RG improved EF potential. A scenario which could have disastrous consequences for the universe.
\begin{figure}
	\begin{center}
		\includegraphics[width=12cm,height=3cm]{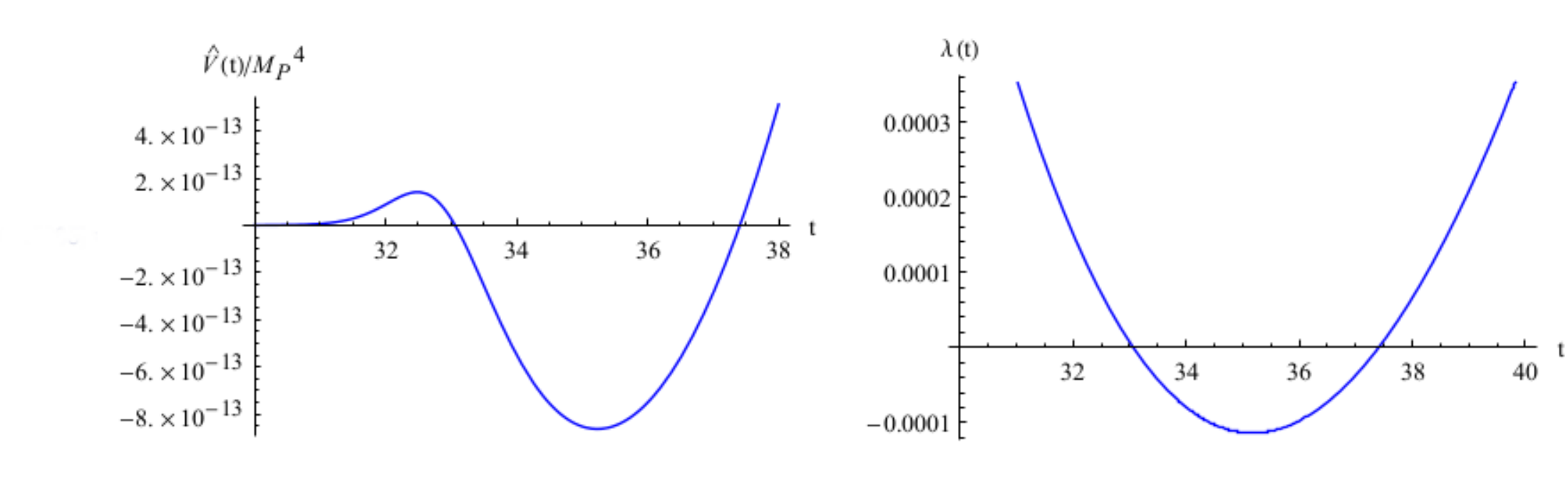}
	\end{center}
	\caption{Left: The RG improved EF potential develops a negative vacuum at high energy scale but below the energy scale of inflation. Right: The running Higgs self-coupling $\lambda$ turns negative for $t_{\mathrm{inst}}<t<t_{\mathrm{inf}}$.}
	\label{FigInst}
\end{figure}

\noindent Various ways to stabilize the EW vacuum have been proposed, among which are thermal effects discussed in \cite{Bezrukov2009b}, additional heavy scalar fields suggested in \cite{Elias-Miro2012}, the inclusion of higher dimensional operators considered in \cite{Branchina2013,Gies2015,Eichhorn2015}, the extended scalaron-Higgs model analysed in \cite{Salvio2015,Kaneda2016,Calmet2016, Wang2017,Ema2017,Ema2017a,He2018,Gorbunov2018,Ghilencea2018,Gundhi2018,Enckell2018,He2019,Ema2019}, or the coupling of a quintessence field to the SM Higgs sector investigated in
\cite{Han2019}.

Besides the division of the parameter space into a ``stable'' regime ($\lambda(t)\geq0$) and a ``non-stable'' regime ($\lambda(t_{\mathrm{inst}})<0$) with $t_{\mathrm{inst}}$ corresponding to an energy scale $E_{\mathrm{inst}}\approx 10^{11}$ GeV, the ``non-stable'' parameter region can be further subdivided into a ``metastable' region and an ``unstable'' region by calculating the (``survival'') probability $\Gamma_{\mathrm{EW}}$ for a quantum tunnelling from the EW vacuum into the negative false vacuum. Comparing the lifetime of the EW vacuum
\begin{align}
\tau_{\mathrm{EW}}\sim\Gamma^{-1}_{\mathrm{EW}},
\end{align}
with the lifetime $\tau_{\mathrm{U}}$ of the universe, metastability (instability) implies $\tau_{\mathrm{EW}}>\tau_{\mathrm{U}}$ ($\tau_{\mathrm{EW}}<\tau_{\mathrm{U}}$). This scenario and its cosmological implications has been studied in many works, see e.g. \cite{Espinosa2008,Elias-Miro2012a,Lebedev2013,Masina2013,Kobakhidze2013,Hook2015,Branchina2015,Herranen2015,Espinosa2015,Burda2015,Burda2016,Kohri2016,Salvio2016a,Joti2017,Salvio2018}.

In contrast to the firm and clear distinction between a ``stable'' region and a ``non-stable'' region in parameter space, which only depends on the values for the SM couplings at the EW scale and the precision of the perturbatively calculated beta functions, the further subdivision of the non-stable region into a ``meta-stable'' and an ``unstable'' region strongly depends on the details of the non-perturbative tunnelling scenario such as e.g. the decay via bubble formation discussed in \cite{Coleman1980} or via the Hawking-Moss instanton proposed in \cite{Hawking1982a}. Summarizing, various tunnelling probabilities have been calculated which, depending on the concrete realization, lead to very different results. Therefore, the conclusion about the ultimate fate of our universe seem to remain obscure.

\subsection{Validity of the effective field theory}
When extrapolating the SM as perturbative quantum field theory up to inflationary energy scales, it is important to ensure that the effective field theory expansion is well under control and does not break down below or at the energy scale of inflation.  
The validity of the effective field theory depends on whether irrelevant higher-dimensional operators are sufficiently suppressed by the associated cutoff $\Lambda$. In the absence of any new physics in between the EW scale and the Planck scale, the natural cutoff in the context of quantum gravity is $\Lambda=\mathrm{M}_{\mathrm{P}}$. However, from a tree-level unitary analysis, it was found in \cite{Burgess2009,Barbon2009,Burgess2010} that the cutoff in non-minimal Higgs inflation is essentially lowered to
\begin{align}
\Lambda=\frac{4\pi M_{\mathrm{P}}}{\xi}.\label{cutoffconst}
\end{align}
In view of the strong non-minimal coupling $\xi\approx 10^4$, the cutoff \eqref{cutoffconst} corresponds to a significantly lower scale than the typical field values during inflation
\begin{align}
\varphi_{\mathrm{inf}}\approx \frac{M_{\mathrm{P}}}{\sqrt{\xi}}.
\end{align}
If true, this would suggest that the predictions based on the low energy approximation \eqref{JFAct} are not valid during inflation, unless an unnatural suppression of higher-dimensional operators is assumed.
However, in the context of small gradient terms and large background fields during inflation, the gravitational interaction strength in the JF parametrization \eqref{JFAct} is not given by $M_{\mathrm{P}}$, but by the effective Planck mass
\begin{align}
M_{\mathrm{P}}^{\mathrm{eff}}(\varphi):=\sqrt{M_{\mathrm{P}}^2+\xi\varphi^2}\geq \sqrt{\xi}\varphi.\label{MPeff}
\end{align}
In \cite{Bezrukov2011, Barvinsky2012} it was shown that the cutoff itself is running $\Lambda\to\Lambda(\varphi)$.
The power counting method of \cite{Burgess2009,Burgess2010}, which lead to \eqref{cutoffconst}, remains valid if the Planck mass $M_{\mathrm{P}}$ is replaced by $M_{\mathrm{P}}^{\mathrm{eff}}$ and leads to a \textit{running} cutoff, which, in view of \eqref{MPeff}, is bounded from below by
\begin{align}
\Lambda(\varphi)=\frac{4\pi\varphi}{\sqrt{\xi}}.\label{cuttoffrunning}
\end{align}
This cutoff, which controls the gradient and curvature expansion, can also be derived from the leading contribution to the quadratic one-loop operator $R^2$ in the large $\xi$ expansion $\sim \xi^2 R^2/(4\pi)^2$ calculated in \cite{Barvinsky2009, Steinwachs2011}.
Comparing this to the tree-level interaction term ${\sim (M_{\mathrm{P}}^2+\xi\varphi^2)R}$, the one-loop contribution is suppressed by a cutoff $\Lambda^2(\varphi)$, which is in agreement with the running cutoff \eqref{cuttoffrunning}.

As demonstrated in \cite{Barvinsky2012}, the gradient and curvature expansion with respect to the running cutoff \eqref{cuttoffrunning} is efficient. Indeed, making use of the JF on-shell relation ${R\sim V/U\sim\lambda\varphi^2/\xi}$, the curvature expansion runs in powers of
\begin{align}
\frac{R}{\Lambda^2(\varphi)}\sim\frac{\lambda}{16\pi^2}\ll 1.\label{curvexp}
\end{align}
The smallness in \eqref{curvexp} follows in the range in which the SM remains perturbative ($\lambda\ll1$). Likewise, as shown in in \cite{Barvinsky2012}, the gradient terms is suppressed even stronger
\begin{align}
\frac{\partial}{\Lambda}\sim\frac{\lambda}{48\pi}\sqrt{\hat{\varepsilon}_{\mathrm{v}}}.\label{gradexp}
\end{align}
According to \eqref{NumberEfolds}, the additional factor $\sqrt{2\hat{\varepsilon}_{v}}$ scales like $1/\hat{N}$ at the beginning of inflation to $\mathcal{O}(1)$ at the end of it. Having established the efficiency of the curvature and gradient expansion \eqref{curvexp} and \eqref{gradexp}, the only higher dimensional operators which could potentially spoil the truncation \eqref{JFAct} of the effective field theory expansion are multi-loop corrections in the form of monomial operators 
\begin{align}
\frac{\varphi^{n}}{\Lambda^{n-4}(\varphi)}.
\end{align}  
While these contributions might be large, they do not spoil the flatness of the inflationary potential as, in view of the running cutoff \eqref{cuttoffrunning}, the ratio $\varphi/\Lambda(\varphi)$ is field independent. 
Moreover, as noted in \cite{Bezrukov2011,Barvinsky2012,Fumagalli2016}, the asymptotic tree-level scale invariance of the model \eqref{JFAct} in the JF is only weakly broken by the quantum logs and leads to an asymptotic shift invariance in the EF, protecting the EF inflaton potential from large quantum corrections. See also \cite{Meissner2007,Shaposhnikov2009,Shaposhnikov2018,Shaposhnikov2018a,Mooij2019,Shaposhnikov2019, Wetterich2019} for a recent discussion on classical and quantum scale (conformal, respectively) invariance. 

\subsection{Quantum cosmology and initial conditions for Higgs inflation}
\label{SecQC}
While the model of non-minimal Higgs inflation provides an answer to the question about the fundamental nature of the inflaton field, which largely remains an open question in many other models of inflation, it does not answer the question about the the initial conditions for inflation or the origin of the universe itself.\footnote{I did not discuss the quantum cosmological implications for Higgs inflation in my talk.}

An attempt to derive the initial conditions for the the inflationary background evolution from quantum cosmology was undertaken in \cite{Barvinsky2010} by reviving the old idea that the universe tunnelled from nothing to existence, see  \cite{Vilenkin1984,Linde1984,Rubakov1984,Zeldovich1984}.
The nucleation process of the universe in the tunnelling scenario is described by a gravitational instanton -- the solution to the Euclidean version of Einstein's equation.
Starting from the Euclidean path integral
\begin{align}
\exp\left(-W\right):=\int\mathcal{D}g\exp\left(-S_{\mathrm{eff}}[g]\right),
\end{align}
the effective gravitational action $S_{\mathrm{eff}}[g]$ is obtained by integrating out all matter fields, collectively denoted by $\psi$,
\begin{align}
\exp\left(-S_{\mathrm{eff}}[g]\right):=\int\mathcal{D}\psi \exp\left(-S[g;\psi]\right)
\end{align}
For the large and slowly varying background fields during inflation, the effective action $S_{\mathrm{eff}}[g]$ admits a local expansion in gradients and curvatures
\begin{align}
S_{\mathrm{eff}}[g]=\frac{M_{\mathrm{P}}^2}{2}\int\mathrm{d}^4x\sqrt{g}\left[2\Lambda_{\mathrm{eff}}-R(g)+\ldots\right].\label{EuclidAct}
\end{align} 
The ellipsis indicate, that we have only kept the leading orders an neglected gradient terms and higher curvature invariants. In the inflationary slow-roll approximation, we identify the quantum effective scalar field potential with an effective cosmological constant 
$M_{\mathrm{P}}^2\Lambda_{\mathrm{eff}}:=V_{\mathrm{eff}}(\varphi)$.
The line element reduces to the Euclidean version of the homogeneous and isotropic FLRW universe
\begin{align}
\mathrm{d}s^2=N(\tau)\mathrm{d}\tau^2+a(\tau)^2\mathrm{d}\Omega^{2}_{(3)},\label{EuclidLine}
\end{align}
with the Euclidean time $\tau$, the volume element of the unit three-sphere $\mathrm{d}\Omega^{2}_{(3)}$, the Euclidean lapse function $N(\tau)$ and the Euclidean scale factor $a(\tau)$. The ``matter'' fields $\psi(\tau,\mathbf{x})$ are associated with all inhomogeneous degrees of freedom, including metric perturbations. On the background \eqref{EuclidLine}, the Euclidean action \eqref{EuclidAct} reduces to
\begin{align}
S_{\mathrm{eff}}[a,N]=12\pi^2M_{\mathrm{P}}^2\int\mathrm{d}\tau N\left[-a+(a')^2a+H^2a^3\right].\label{ActEuclSym}
\end{align}
In \eqref{ActEuclSym} we have defined $a':=N^{-1}\mathrm{d}/\mathrm{d}\tau$. The tunnelling instanton is obtained for the gauge choice $N=-1$ as stationary configuration with respect to variations of the lapse function in the saddle point approximation
\begin{align}
\left(a'\right)^2=1-H^2a^2.\label{EuclEOM}
\end{align}
The Euclidean Friedman equation \eqref{EuclEOM} has a turning point at $a_{+}$ corresponding to the equator of the the Euclidean $S^4$ half-sphere. The positive solution of \eqref{EuclEOM} is
\begin{align}
a(\tau)=H^{-1}\sin\left(H\tau\right),\label{sol}
\end{align}
where we have fixed the constant of integration by the condition $\mathrm{d}a/\mathrm{d}\tau|_{a=0}=1$, which follows from \eqref{EuclEOM} at $a_{-}=0$. The tunnelling probability distribution function (PDF) for $H^2=\Lambda_{\mathrm{eff}}/3$ and $\Lambda_{\mathrm{eff}}>0$ is obtained from \eqref{ActEuclSym} by integrating the configurations \eqref{sol} from the pole of the Euclidean half-sphere at $a_{-}=a(\tau_{-})=0$ to the nucleation point $a_{+}=a(\tau_{+})=1/H$ at the equator \footnote{Note that \eqref{InstTun} coincides with the PDF obtained from the semiclassical expansion of the Wheeler-DeWitt equation ${e^{-W}=|\Psi|^2}$ with the tunnelling wave function of the universe $\Psi$. Corrections from canonical quantum gravity to the inflationary power spectra have been first derived for the minimally coupled theory \cite{Kiefer2012a} and later for the general scalar-tensor theory in \cite{Steinwachs2018,Steinwachs2019}.} 
\begin{align}
P(\varphi)=\exp\left[-W(\varphi)\right]=\exp\left[-\frac{24\pi^2M_{\mathrm{P}}^4}{V_{\mathrm{eff}}(\varphi)}\right].\label{InstTun}
\end{align}
At the moment of nucleation $\tau_{+}=\pi/2H$, the solution \eqref{sol} can be analytically continued $\tau\to i t$ to the Lorentzian regime
\begin{align}
a_{\mathrm{L}}(t)=\frac{1}{H}\cosh\left(Ht\right).\label{scaleLor}
\end{align}
The tunnelling instanton \eqref{InstTun} can be interpreted as representing the PDF of scale factors \eqref{scaleLor} in the quantum ensemble of de Sitter models after nucleation, i.e. the realizations of scale factors $a_{L}$ (with different $H(\varphi)=\sqrt{\Lambda_{\mathrm{eff}}(\varphi)/3}$) are distributed according to \eqref{InstTun}. 
The maximum $\varphi_{\mathrm{max}}$ of the potential $V_{\mathrm{eff}}(\varphi)$ corresponds to a peak in the distribution \eqref{InstTun} and can be interpreted as the most probable value for the inflationary trajectory to start, i.e. provides the initial condition for inflation

We note that the quantum cosmological scenario could in principle be falsified by observations, as it must satisfy a consistency condition -- the value $\varphi_{\mathrm{max}}$ determined by the tunnelling scenario has to be compatible with the value $\varphi_{\mathrm{inf}}$ derived from the energy scale of inflation, i.e. $\varphi_{\mathrm{max}}\geq\varphi_{\mathrm{inf}}$.
The energy scale of inflation could be inferred from the detection of primordial gravitational waves,
\begin{align}
E^{\mathrm{obs}}_{\mathrm{inf}}=M_{\mathrm{P}}\left(\frac{3}{2}\pi^2 A_{\mathrm{t},*}\right)^{1/4}.
\end{align}
The observational data \eqref{CMBAmpl}, \eqref{AsCMB} and \eqref{Planckr} imply the upper bound $E_{\mathrm{obs}}^{\mathrm{inf}}\lesssim 10^{16}$ GeV.

When this general formalism is applied to the model of non-minimal Higgs inflation, the RG improved EF potential \eqref{EFPot} enters the tunnelling PDF \eqref{InstTun},
\begin{align}
P(\varphi)\simeq\exp\left[-96\pi^2\frac{\xi^2}{\lambda}\left(1+\frac{2M_{\mathrm{P}}^2}{\xi \varphi^2}\right)\right].\label{PDF}
\end{align}
Here, the couplings $\lambda(t)$ and $\xi(t)$ are function of the logarithmic RG scale ${t=\log(\varphi/M_{\mathrm{t}})}$ and we have neglected the wave function renormalization of the Higgs boson, i.e. set $Z=1$. The distribution \eqref{PDF} has a sharply peaked maximum at
\begin{align}
\varphi_{\mathrm{max}}^2=-\left.\frac{64\pi^2M_{\mathrm{P}}^2}{\xi \mathbf{A}_{\mathrm{I}}}\right|_{t_\mathrm{max}}.\label{phimax}
\end{align}
Here, $\mathbf{A}_{\mathrm{I}}$ is the inflationary anomalous scaling defined in \eqref{InfAnDim}.
The value \eqref{phimax} for $\varphi_{\mathrm{max}}$ satisfies the consistency condition $\varphi_{\mathrm{\max}}\geq\varphi_{\mathrm{inf}}$ as the value $\varphi_{\mathrm{inf}}$ at horizon crossing was found to be related to $\varphi_{\mathrm{max}}$ by
\begin{align}
\frac{\varphi^2_{\mathrm{inf}}}{\varphi^2_{\mathrm{max}}}=1-\exp\left(-\hat{N}\frac{\mathbf{A}_{\mathrm{I}}(t_{\mathrm{end}})}{48\pi^2}\right).
\end{align} 
Thus for wavelengths longer than the pivotal one, the moment of horizon crossing $\varphi_{\mathrm{inf}}$ comes closer to the moment of nucleation $\varphi_{\mathrm{max}}$ but always chronologically stays behind it $\varphi_{\mathrm{max}}>\varphi_{\mathrm{inf}}$ and approaches it in the limit $N\to\infty$.

The quantum cosmological analysis provides a complete picture as it suggests that non-minimal Higgs inflation might even predict its own initial conditions in a self-consistent way, followed by a successful inflationary phase and a subsequent transition to the SM at low energies.  

\section{Quantum field parametrization dependence in cosmology}
\label{QFP}
In this section, I discuss another aspect of Higgs inflation, which is connected to a more general field theoretical problem. The predictions \eqref{ObsNMHI} were derived by transforming between the JF parametrization \eqref{JFAct} and the EF parametrization \eqref{EFACT}. In fact, the class of models leading to inflationary predictions equivalent to those of non-minimal Higgs inflation \eqref{ObsNMHI} is much larger and also includes geometrical modifications of general relativity.  

\subsection{Starobinsky inflation}
The first model of inflation was proposed in \cite{Starobinsky1980}, which according to recent data \cite{Akrami2018} is still one of the most favoured models, see also the analysis in \cite{Martin2014}. In addition to the Einstein-Hilbert action, which is linear in the scalar curvature, the action of Starobinsky's model includes the square of the Ricci scalar 
\begin{align}
S[g]=\frac{M_{\mathrm{P}}^2}{2}\int\mathrm{d}^4x\sqrt{-g}\left[R+\frac{R^2}{6M^2}\right],\label{fRAct}
\end{align}
Beside the Planck mass $M_{\mathrm{P}}$, the mass scale $M$ is the only new scale in the model. Due to the fourth-order derivatives implicit in the $R^2$ term, the theory effectively propagates a massive scalar particle -- the \textit{scalaron}. In contrast to other higher derivative modifications of Einstein's theory, which involve quadratic invariants build from the Ricci tensor and the Riemann tensor, $f(R)$ gravity does not suffer from the Ostrogradski instability and the associated problem of the higher derivative massive spin-two ghost, which spoil the unitarity of the corresponding quantum theory as discussed in \cite{Stelle1977} and \cite{Woodard2009}.
The additional scalar degree of freedom can be made explicit by transforming the action \eqref{fRAct} into its scalar-tensor representation. Performing the field redefinitions
\begin{align}
\hat{g}_{\mu\nu}=\exp\left(\sqrt{\frac{2}{3}}\frac{\hat{\chi}}{M_{\mathrm{P}}}\right)g_{\mu\nu},\qquad\hat{\chi}(R)=\sqrt{\frac{2}{3}}M_{\mathrm{P}}\ln\left(1+\frac{R}{3M}\right),\label{TrafofR}
\end{align}
the scalaron $\hat{\chi}$ becomes manifest in the EF formulation
\begin{align}
S[\hat{g},\hat{\chi}]=\int\mathrm{d}^4x\sqrt{-\hat{g}}\left[\frac{M_{\mathrm{P}}^2}{2}\hat{R}-\frac{1}{2}\hat{g}^{\mu\nu}\partial_{\mu}\hat{\chi}\partial_{\nu}\hat{\chi}-\hat{V}(\hat{\chi})\right],
\end{align}
with the Starobinsky EF potential
\begin{align}
\hat{V}(\hat{\chi})=\frac{3}{4}M_{\mathrm{P}}^2M^2\left(1-e^{-\sqrt{\frac{2}{3}}\frac{\hat{\chi}}{M_{\mathrm{P}}}}\right)^2.\label{StarPot}
\end{align}
Evaluating \eqref{CMBAmpl}-\eqref{TTS} for \eqref{StarPot} to leading order in $N_{*}\gg1$ results in
\begin{align}
A_{\mathrm{s}}^{*}=\frac{N_{*}^2}{24 \pi^2 }\frac{M^2 }{M_{\mathrm{P}}^2 },\qquad n_{\mathrm{s}}^{*}=1-\frac{2}{N_{*}}\approx0.9667,\qquad r^{*}=\frac{12}{N_{*}^2}\approx 0.0033,\label{ObsStar}
\end{align}
where the numerical values are again obtained for $N_{*}=60$.
The predictions for $n_{\mathrm{s}^{*}}$ and $r^{*}$ are not only in excellent agreement with the current observational bounds \eqref{Planckns} and \eqref{Planckr}, they are identical to the tree-level predictions of non-minimal Higgs inflation \eqref{ObsNMHI}.
The normalization condition \eqref{AsCMB} fixes the only free parameter ${M\approx 10^{-5}M_{\mathrm{P}}}$. Identifying $M=\sqrt{\lambda/3\xi^2}M_{\mathrm{P}}$ shows that non-minimal Higgs inflation and Starobinsky inflation are indistinguishable regarding their inflationary observables.
However, the couplings to other fields, the predictions for reheating and the inclusion of quantum corrections makes them at least in principle observationally distinguishable \cite{Bezrukov2012}. The degenerate inflationary predictions of these two models can be explained naturally, as both are part of a common universality class of inflationary models, see \cite{Mukhanov2013}.

Before discussing in more detail the general question of equivalence between different field parametrizations, another interesting point regarding the field transformations \eqref{TrafoJFEF} and \eqref{TrafofR} is the following:
In the geometric formulation of Starobinsky's model, the term quadratic $R^2$ dominates over the linear term for high energies $R/6M^2\gg1$ and leads to an asymptotic scale invariance of the action \eqref{fRAct}. Likewise, for high energies $\varphi\gg M_{\mathrm{P}}/\sqrt{\xi}$, the non-minimally coupled term dominates and leads to the aforementioned asymptotic scale invariance of the action \eqref{JFAct}. In both cases, inflation is realized for the approximate scale-invariant quasi de Sitter phase and ended when the scale invariance is broken by the Einstein-Hilbert term. Moreover, the transformations \eqref{TrafoJFEF} and \eqref{TrafofR} show that this asymptotic scale invariance 
\begin{align}
g_{\mu\nu}\to g'_{\mu\nu}\alpha^{-2}g_{\mu\nu},\qquad \varphi\to\varphi'=\alpha \varphi,
\end{align}
translates into an asymptotic shift symmetry of the scalar field in the EF formulation
\begin{align}
\hat{\varphi}\to\hat{\varphi}'=\hat{\varphi}+\ln\alpha,
\end{align}
with constant scaling parameter $\alpha$. An approximate shift symmetry, in turn, naturally explains why the inflationary quasi de Sitter phase is realized by an almost flat, quasi-constant potential \eqref{EFPot} and \eqref{StarPot} in the EF formulation. 

\subsection{Classical and quantum equivalence in cosmology}
\label{Eq}
The equivalence between the JF and EF formulations in non-minimal Higgs inflation as well as the equivalence between non-minimal Higgs inflation and Starobinsky inflation are just particular realizations of a more general equivalence. In fact, the gravity-Higgs sector \eqref{JFAct} is a particular case of the more general scalar-tensor theory
\begin{align}
S[g,\varphi]=\int\mathrm {d}^4x\sqrt{-g}\left[U(\varphi)R-\frac{1}{2}G(\varphi)\partial_{\mu}\varphi\partial^{\mu}\varphi-V(\varphi)\right]\label{ActST}
\end{align}
with general functions $U(\varphi)$, $G(\varphi)$ and $V(\varphi)$ which parametrize the non-minimal coupling to gravity, a non-standard kinetic term and an arbitrary potential. The action \eqref{ActST} covers almost all single-field model of inflation.
Likewise, Starobinsky's model \eqref{fRAct} is just a specific case of the more general class of $f(R)$ models
\begin{align}
S[g]=\int\mathrm{d}^4x\sqrt{-g}f(R),\label{FRACT}
\end{align}
with an arbitrary function $f$.
Note that since $f(R)$ is a function of the undifferentiated Ricci scalar $R$, apart from the scalaron, there are no additional propagating degrees of freedom despite the arbitrariness of $f$.

Provided $U\neq0$, the transition of the general scalar-tensor theory \eqref{ActST} to the EF action \eqref{EFACT} is achieved by performing the field redefinitions
with the EF potential
\begin{align}
\hat{g}_{\mu\nu}=\frac{2U}{M_{\mathrm{P}}^2}g_{\mu\nu},\qquad \left(\frac{\partial\hat{\varphi}}{{\partial\varphi}}\right)^2=\frac{M_{\mathrm{P}}^2}{2}\frac{GU+3U_{,\varphi}^2}{U^2},\qquad\hat{V}(\hat{\varphi})=\left(\frac{M_{\mathrm{P}}^2}{2}\right)^2\frac{V(\hat{\varphi})}{U^2(\hat{\varphi})}.\label{STEF}
\end{align}
Similarly, provided $f_{,R}\neq0$ and $f_{,RR}\neq0$, the $f(R)$ theory \eqref{FRACT} is mapped to the EF action \eqref{EFACT} by performing the field redefinitions
with the EF potential
\begin{align}
\hat{g}_{\mu\nu}=\frac{2f_{,R}}{M_{\mathrm{P}}^2}g_{\mu\nu},\qquad\hat{\varphi}(R) = M_{\mathrm{P}}\sqrt{\frac{3}{2}}\ln f_{,R},\qquad \hat{V}(R)=\left(\frac{M_{\mathrm{P}}^2}{2}\right)^2\frac{Rf_{,R}-f}{f_{,R}^2}.\label{fREF}
\end{align}
Comparing the transformations \eqref{STEF} and \eqref{fREF} implies the identifications 
\begin{align}
U\leftrightarrow f_{,R},\qquad V\leftrightarrow Rf_{R}-f,\qquad GU+3U_{,\varphi}^2\leftrightarrow 3f_{,RR}.
\end{align}
At the level of the classical action, the classical equations of motion and the linear perturbations propagating on the background (which is a solution of the classical equations of motion), the formulations in different parametrizations related by non-singular but arbitrary non-linear transformations are mathematically equivalent. All parametrizations correspond to one and the same physical theory, just expressed in terms of different field variables.
In contrast to the non-linear but ultra-local transformation \eqref{STEF} relating the JF and EF parametrizations of the scalar-tensor theory, the transformations \eqref{fREF} relating the $f(R)$ theory with the EF scalar-tensor parametrization involves in addition derivatives -- reflecting the presence of the additional propagating scalar degree of freedom. 

\subsection{Quantum equivalence and renormalization}
In Sec. \eqref{Eq}, the identical inflationary predictions of the model of non-minimal Higgs inflation \eqref{JFAct} and Starobinsky's model of inflation \eqref{fRAct} was traced back to the more general classical equivalence between scalar-tensor theories \eqref{ActST} and $f(R)$ gravity \eqref{FRACT}. In this section, I discuss the equivalence between different field parametrizations at the quantum level.

As emphasized in Sec. \ref{QC}, quantum corrections and the RG improvement are crucial.
In the RG improved treatment, the inflationary observables are expressed in terms of the running couplings evaluated at the energy scale of inflation. The running couplings are solutions to the RG equations. The RG system is determined by the beta functions. The beta functions are derived order-by-order in perturbation theory from the ultraviolet divergences of the theory, or, in non-perturbative approaches, such as the asymptotic safety program proposed in \cite{Weinberg1980}, from solving the Wetterich equation for the functional RG flow of the averaged effective action within a given truncation \cite{Wetterich1993}. 

\begin{figure}[h!]
	\begin{center}
		\includegraphics[width=9cm,height=3cm]{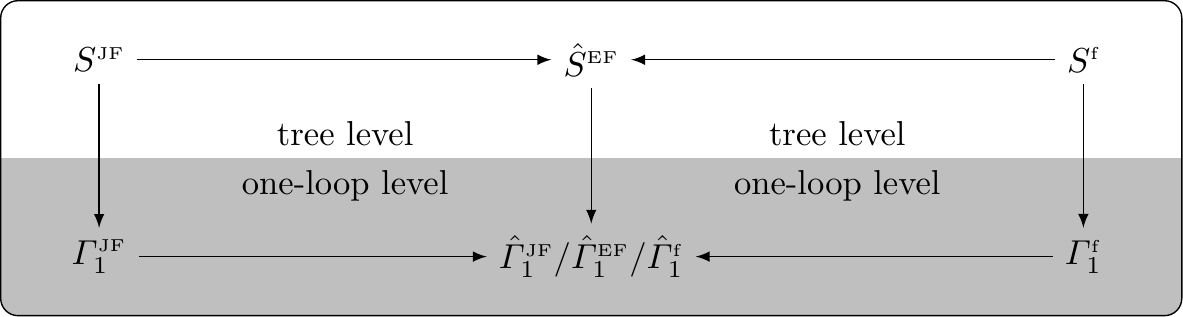}
	\end{center}
	\caption{Transition between formulations in terms of different field parametrizations at the classical and quantum level. If the classical equivalence extends to the one-loop quantum level, the diagram must commute.}
	\label{FrameFig}
\end{figure}

In the context of the field transformation, relating different representations of the same classical theory, an important question arises: does the RG improvement derived in different field parametrizations lead to the same results for the inflationary observables, or, in other words, does the classical equivalence between the different formulations extend to the quantum level?
In order to answer this question to leading order in the perturbative loop expansion, we explicitly calculate the one-loop divergences in different parametrizations and compare the results. Phrased diagrammatically, the question of quantum  equivalence reduces to the question of whether the diagram shown in Fig. \ref{FrameFig} commutes or not.

The combination of the background field method with heat kernel techniques provides an efficient and manifest covariant tool to calculate the ultraviolet divergences in curved spacetime. The Euclidean quantum effective action $\Gamma[\phi_0]$ for a theory with bare action functional $S[\phi]$ and generalized field $\phi^{i}$ can be defined by the functional integro-differential equation\footnote{The compact DeWitt notation combines the collection of discrete bundle indices $A,B,\ldots$ with the continuous spacetime arguments $x,y,\dots$ into the generalized DeWitt index $i=(A,x)$,  where it is understood that the Einstein summation convention for the DeWitt indices $i,j\ldots$ includes integration over the continuous spacetime arguments, i.e. $\phi^{i}\phi_i=\int\mathrm{d}^4x\phi^{A}(x)\phi_{A}(x)$.}
\begin{align}
\exp\left(-\Gamma\right)=\int\mathcal{D}\phi{\mathcal M}(\phi)\exp\left\{-S[\phi]-(\phi^{i}_{0}-\phi^{i})\Gamma_{,i}[\phi_0]\right\},\label{defeq}
\end{align}
Here, $\phi_0^{i}$ is the mean field (one-point correlation function) and ${\mathcal M}(\phi)$ is the functional measure. 
Expanding the action $S[\phi]$ around the mean field $\phi_0^{i}$,
this equation serves as starting point for a perturbative expansion of $\Gamma$ in powers of $\hbar$, corresponding to the number of closed loops in terms of a Feynman diagrammatic representation
\begin{align}
\Gamma[\phi_0]=S[\phi_0]-\hbar\ln \mathcal{M}(\phi_0)+\hbar \Gamma_{1}[\phi_0]+\mathcal{O}\big(\hbar^2\big).
\end{align}
Neglecting the contribution from the (ultra) local measure $\mathcal{M}(\phi_0)\propto1+\delta(0)(\ldots)$, the one-loop effective action $\Gamma_{1}$ is given in terms of the functional trace
\begin{align}
\Gamma_{1}=\frac{1}{2}\mathrm{Tr}\ln F_{ij},
\end{align}
with the fluctuation operator defined by the Hessian of the classical action
\begin{align} 
F_{ij}(\nabla):=\partial_{i}\partial_jS=\left.\frac{\delta^2 S[\phi]}{\delta\phi^{i}\delta\phi^{j}}\right|_{\phi=\bar{\phi}}.\label{FlucOp}
\end{align}
At the level of the one-loop approximation, the background field  $\bar{\phi}^{i}$ might be identified with the mean field $\phi_{0}^{i}$.\footnote{A systematic order-by-order renormalization procedure for arbitrary loop order, which ensures the gauge invariant structure of the counterterms and keeps tract of the background field and mean field separately is proposed in \cite{Barvinsky2018}.}.
The one-loop beta functions are proportional to the counterterms, which in the $\overline{\mathrm{MS}}$ scheme are determined by $-\Gamma^{\mathrm{div}}_{1}$. For minimal second-order fluctuation operators
\begin{align}
\mathbf{F}(\nabla)=\mathbf{1}\Delta+\mathbf{\Pi},\label{MinSecOp}
\end{align}
the Schwinger-DeWitt technique originally developed in \cite{DeWitt1965} allows to calculate the one-loop divergences in a closed form.
Here, bundle indices are suppressed and operator valued quantities such as e.g. $\mathbf{F}=F^{A}_{\;B}=\mathcal{G}^{AC}F_{CB}$ are written in bold face, $\Delta=-g^{\mu\nu}\nabla_{\nu}\nabla_{\nu}$ is the positive definite Laplacian and $\mathbf{\Pi}$ is the potential part acting multiplicatively. 
The one-loop UV divergences in dimensional regularization $d=4-2\varepsilon$ are isolated as poles $1/\varepsilon$ in the limit $\varepsilon\to0$. For the operator \eqref{MinSecOp}, the divergent part of the one-loop effective action in curved spacetime reads
\begin{align}
\Gamma_{1}^{\mathrm{div}}=-\frac{1}{32\pi^2\varepsilon}\int\mathrm{d}^4x\sqrt{g}\,\mathrm{tr}\,\mathbf{a}_2(x,x).
\end{align}
It involves the bundle trace $\mathrm{tr}$ of the coincidence limit of the second Schwinger-DeWitt coefficient, which up to total derivative terms is given by
\begin{align}
\mathbf{a}_{2}(x,x)=\frac{1}{180}\left(R_{\mu\nu\rho\sigma}R^{\mu\nu\rho\sigma}-R_{\mu\nu}R^{\mu\nu}\right)\mathbf{1}+\frac{1}{2}\mathcal{R}_{\mu\nu}\mathcal{R}^{\mu\nu}+\frac{1}{2}\left(\mathbf{\Pi}-\frac{R}{6}\mathbf{1}\right)^2.
\end{align}
The bundle curvature (suppressing bundle indices) is defined as ${\mathcal{R}_{\mu\nu}:=[\nabla_{\mu},\nabla_{\nu}]\mathbf{\phi}}$.
For higher-order and non-minimal operators $\mathbf{F}$, the Schwinger-DeWitt algorithm is not directly applicable and more general techniques developed in \cite{Barvinsky1985} are required. For operators with a degenerate principal part or with an effective Laplacian, even these generalized techniques cannot be applied directly and other methods discussed in \cite{Ruf2018,Ruf2018b,Ruf2018c,Heisenberg2019} must be employed. 

In order to perform the explicit comparison illustrated in Fig. \ref{FrameFig}, first the individual calculations in the different field parametrizations have to be performed. The derivation of the one-loop divergences for the EF action \eqref{EFACT} with arbitrary EF potential $\hat{V}$ has been performed in the EF field parametrization in \cite{Barvinsky1993,Steinwachs2011}. The corresponding calculation for the general scalar-tensor action \eqref{ActST} in the JF field parametrization has been carried out in \cite{Shapiro1995} and has been generalized in \cite{Steinwachs2011} for a $O(N)$ symmetric multiplet of scalar fields. Finally, the one-loop divergences for the $f(R)$ action were obtained in \cite{Ruf2018}. Note that in all cases the off-shell divergences were calculated on a general background, which is crucial in order to uniquely ascribe the individual coefficients to different operator structures.\footnote{On particular symmetric backgrounds such as e.g. spaces of constant curvature, the different curvature invariants degenerate and their individual contributions are no longer resolvable.} 

Using the transition formulas \eqref{STEF} and \eqref{fREF}, the one-loop results for the JF scalar-tensor theory and for $f(R)$ gravity were transformed back to the EF and compared with the one-loop divergences directly obtained in the EF to explicitly check the commutativity of the diagram Fig. \ref{FrameFig}. The comparison between the EF and JF parametrizations has been performed in \cite{Kamenshchik2015}, while the comparison between the $f(R)$ and scalar-tensor formulations has been carried out in \cite{Ruf2018b}. 
In both cases, the one-loop comparison showed that the divergent part of the off-shell one-loop effective action, calculated in different parametrizations, does not coincide, implying that the classical equivalence is lost at the quantum level. However, in both cases it was also found that the quantum equivalence is restored for the on-shell reduction. The on-shell equivalence is in agreement with formal S-matrix equivalence theorems formulated in \cite{Chisholm1961, Kamefuchi1961,Coleman1969,Kallosh1973}.\footnote{Note however, that some of the propositions in the formal theorems, such as e.g. the assumption of asymptotically free states, are in general not satisfied in the context of gravity and cosmology.}

A naive application of the of the RG improvement by replacing the coupling constants in cosmological observables by the running constants leads to ambiguous results. The couplings are solutions of the system of RG equations which are in turn controlled by the off-shell beta function. Since the off-shell beta functions are derived from the off-shell divergences, they inherit the off-shell parametrization ambiguity. 

\subsection{Geometry of field space and field covariant formalism}
From a quantum field theoretical point of view, the configuration space of fields (including the spacetime metric) might formally be viewed as a differentiable manifold. In this geometric setup, different field parametrizations simply correspond to different local coordinates and the quantum off-shell parametrization dependence is revealed as a failure of the non-field covariant mathematical formalism underlying the ordinary definition of the quantum effective action.
In more detail, the analysis shows that the ordinary quantum effective action is not a true scalar with respect to configurations space diffeomorphisms (although it is of course a scalar with respect to spacetime diffeomorphisms). 
Within a field covariant formalism, it becomes meaningless to talk about a preferred physical field parametrization -- any frame is as good as any other. 
In this sense, the off-shell quantum frame ambiguity finds a natural resolution in terms of the geometrically defined effective action, introduced in \cite{Vilkovisky1984}. Instead of trying to select a preferred parametrization on physical grounds, the quantum effective action is defined in a manifestly field parametrization invariant way. 
The fact that the ordinary effective action is not invariant under field reparametrizations can eb traced back to the geometrically meaningless coordinate difference $(\phi^{i}_0-\phi^{i})$ which enters in the exponent of the defining equation \eqref{defeq}. According to Vilkovisky, it should be replaced by the geometrical meaningful two-point quantity $\sigma^{i}(\phi_{0},\phi)$,
\begin{align}
(\phi^{i}_0-\phi^{i})\to\sigma^{i}(\phi_{0},\phi):=\mathcal{G}^{ij}\nabla_{j}\sigma(\phi_{0},\phi).\label{sigma}
\end{align}
The bi-scalar $\sigma(\phi_0,\phi)$ is Synge's world function on the configuration space manifold.
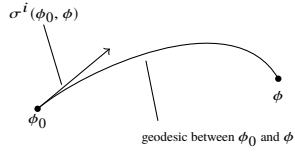
\begin{figure}
	\begin{center}
		\begin{tikzpicture}[scale=1.6,xshift=-3cm]
		\begin{scope}[xshift=-4cm]
		\begin{scope} [xshift=-1.8cm]  
		\draw (1,0) .. controls +(40:0.5cm) and +(120:0.75cm) .. (3,+0.25);
		\draw[black](2,-0.1)--(1.9,0.4);
		\node[black](a) at(2.5,-0.25){\tiny geodesic between $\phi_0$ and $\phi$};
		\draw[black](1.2,0.2)--(1.05,0.7);
		\node[black](a) at(1.05,0.8){\tiny $\sigma^{i}(\phi_0,\phi)$};
		\node[black](a) at(1,-0.1){\tiny $\phi_0$};
		\node[black](a) at(3,0.1){\tiny $\phi$};
		\draw[color=black,thin,->](1,0.0)--(1.6,0.5);
		\fill[color=black] (1,0) circle (0.75pt);
		\fill[color=black] (3,+0.25) circle (0.75pt);
		\end{scope}
		\end{scope}
		\end{tikzpicture}
	\end{center}
	\caption{Synge's world function is a measure of the (squared) geodesic distance between the two points $\phi_0$ and $\phi$ in configuration space.}
	\label{Synge}
\end{figure}

\noindent Here, $\mathcal{G}^{ij}(\phi)$ is the inverse of the configuration space metric $\mathcal{G}_{ij}(\phi)$ and $\nabla_{i}$ is the field covariant derivative, which defines the configuration connection $\Gamma^{k}_{ij}(\phi)$. The derivative \eqref{sigma} is a vector at $\phi_{0}$ and a scalar at $\phi$ as depicted in Fig. \ref{Synge}. In the one-loop approximation, the identification of the functional measure with ${\mathcal{M}=\sqrt{\mathrm{Det}(\mathcal{G}_{ij})}}$ and the replacement of the coordinate difference ${(\phi^{i}_0-\phi^{i})\to \sigma^{i}(\phi_0,\phi)}$ in in \eqref{defeq} , leads to a replacement of partial derivatives by covariant ones in the definition of the fluctuation operator \eqref{FlucOp}, 
\begin{align}
\partial_{i}\partial_{j}S\to \nabla_{i}\nabla_{j}S=\partial_{i}\partial_{j}S-\Gamma_{ij}^{k}\partial_{k}S.\label{covS}
\end{align}
Independently of the concrete prescription for the  explicit construction of the configuration space metric $\mathcal{G}_{ij}$ and the Vilkovisky connection $\Gamma^{i}_{jk}$, it is clear that the additional term in the covariant construction \eqref{covS} is proportional to $\partial_{i}S$ and vanishes on-shell $\partial_{i}S=0$. Hence, on-shell the one-loop approximation of the covariant Vilkovisky-DeWitt effective action reduces to the one-loop approximation of the ordinary effective action. 

As first discussed in \cite{Steinwachs2013, Steinwachs2014, Kamenshchik2015}, the problem of quantum frame dependence in cosmology is just a particular manifestation of this more general field theoretic problem of field parametrization dependence and the construction of a unique off-shell extension of the quantum effective action. For related work on classical and quantum frame dependence in cosmology, see also \cite{Magnano1994,Capozziello1997,Faraoni1999,Nojiri2001,Flanagan2004,Deruelle2011,Calmet2013,Prokopec2013,Chiba2013,Postma2014,Moss2014a,Domenech2015,Jarv2015,Kamenshchik2016,Herrero-Valea2016,Burns2016,Jarv2017,Karam2017,Bhattacharya2017,Bounakis2018,Karamitsos2018, Ohta2018,Falls2019,Finn2019}.   
However, besides the technical difficulties in the explicit construction of the Vilkovisky-DeWitt effective action, it is questionable whether such a construction is actually required in the context of cosmological observables. A direct construction of manifestly reparametrization and gauge invariant cosmological observables  ($n$-point correlation functions) would most likely be more efficient and would also lead to unique results. 

\section{Conclusions}
\label{Conclusions}
In this contribution, I reviewed and discussed various classical and quantum aspects of the model of non-minimal Higgs inflation. Summarizing, this model offers a theoretically well motivated and phenomenologically successful unified description of particle physics with inflationary cosmology. The identification of the inflaton field with the SM Higgs boson provides a fundamental explanation of the scalar field which drives the accelerated expansion of the early universe. The RG flow connects the EW scale with the energy scale of inflation and supports the scenario that the SM could be a perturbative quantum field theory all the way up to the Planck scale.
The effective field theory expansion of the model can be arranged in a controlled way upon the introduction of a field dependent cutoff.
Moreover, within the quantum cosmological tunnelling scenario, the models predicts its own initial conditions for the onset of inflation in a self-consistent way.

In spite of all these appealing properties, I also highlighted several questions and open problems.
For the central values of the SM Higgs boson mass and the Yukawa top quark mass as measured by collider experiments at the EW scale, the RG flow of the SM drives the quartic Higgs self-coupling to negative values, resulting in the formation of a negative vacuum at high energy scales, which in turn triggers an instability of the EW vacuum. This instability has to be cured, either by finding ways to directly stabilize the RG improved Higgs potential or by invoking a mechanism which prevents the EW vacuum to decay within the lifetime of our universe.
      
Besides, I discussed the ambiguity in the definition of the perturbative off-shell beta function which results from a quantization in different field parametrizations related by a Weyl transformation of the metric field and a highly non-linear transformation of the scalar field. 
I also illustrated the classical equivalence between scalar-tensor sector of non-minimal Higgs inflation and Starobinsky's model of inflation, which corresponds to a subclass of geometric $f(R)$ modifications of General Relativity. I discussed how, within the ordinary definition of the quantum effective action, off-shell quantum corrections break this equivalence and how this equivalence is restored on-shell. I gave a natural explanation of these results in terms of Vilkovisky's geometrically defined parametrization invariant off-shell extension of the quantum effective action and emphasized the importance of and the need for gauge and parametrization invariant cosmological observables. Finally, coming back to Hermann Weyl, I hope that already from the very limited discussion in this contribution, it became evident how his original ideas introduced 100 years ago, still strongly influence active research in theoretical physics today.

\begin{acknowledgement}
It is a pleasure to thank Silvia De Bianchi and Claus Kiefer for inviting me to this very stimulating interdisciplinary conference and the physics centre in Bad Honnef for the warm hospitality. I also thank the other participants for many interesting after-dinner discussions. 
\end{acknowledgement}

\bibliography{HiggsFieldInCosmologyRefV2}
\bibliographystyle{harvard} 
\end{document}